\documentclass[journal=apchd5, manuscript=article]{achemso}

\usepackage{xr-hyper}
\setkeys{acs}{maxauthors=10}
\setkeys{acs}{etalmode=truncate}

\makeatletter
\newcommand*{\addFileDependency}[1]{
  \typeout{(#1)}
  \@addtofilelist{#1}
  \IfFileExists{#1}{}{\typeout{No file #1.}}
}
\makeatother

\makeatletter
\newcommand*{\forcekeywords}{
  \acs@keywords@print
  \let\acs@keywords@print\relax
}
\makeatother

\newcommand*{\myexternaldocument}[1]{%
    \externaldocument{#1}%
    \addFileDependency{#1.tex}%
    \addFileDependency{#1.aux}%
}

\newcommand{\onlinecite}[1]{\hspace{-1 ex} \nocite{#1}\citenum{#1}}

\myexternaldocument{SI}

\usepackage[version=3]{mhchem} 
\usepackage{chemformula} 
\usepackage[T1]{fontenc} 

\usepackage{siunitx}
\usepackage{mathrsfs, mathtools}
\usepackage{xcolor}
\usepackage{amsmath, amssymb, gensymb, textcomp}
\usepackage{multirow, array}
\usepackage{graphicx}
\usepackage[font=small,labelfont=bf]{caption}
\usepackage{wrapfig}
\usepackage{multirow}
\usepackage{placeins}
\usepackage{float}
\DeclareSIUnit\angstrom{\text {Å}}

\author{Tanay Paul}
\altaffiliation{These authors contributed equally to this work.}
\affiliation[UTChemE]{McKetta Department of Chemical Engineering, University of Texas at Austin, Austin, Texas 78712, United States}
\author{Allison M. Green}
\altaffiliation{These authors contributed equally to this work.}
\affiliation[UTChemE]{McKetta Department of Chemical Engineering, University of Texas at Austin, Austin, Texas 78712, United States}
\author{Delia J. Milliron}
\affiliation[UTChemE]{McKetta Department of Chemical Engineering, University of Texas at Austin, Austin, Texas 78712, United States}
\alsoaffiliation[UTChem]{Department of Chemistry, University of Texas at Austin, Austin, Texas 78712, United States}
\email{milliron@che.utexas.edu}
\author{Thomas~M.~Truskett}
\affiliation[UTChemE]{McKetta Department of Chemical Engineering, University of Texas at Austin, Austin, Texas 78712, United States}
\alsoaffiliation[UTPhys]{Department of Physics, University of Texas at Austin, Austin, Texas 78712, United States}
\email{truskett@che.utexas.edu}

\title[An \textsf{achemso} demo]
  {Plasmon Polaritons in Disordered Nanoparticle Assemblies}

\abbreviations{IR,NMR,UV}

\begin{document}

\begin{abstract}
Multilayer assemblies of metal nanoparticles can act as photonic structures, where collective plasmon resonances hybridize with cavity modes to create plasmon-polariton states. For sufficiently strong coupling, plasmon polaritons qualitatively alter the optical properties of light-matter systems, with applications ranging from sensing to solar energy. However, results from experimental studies have raised questions about the role of nanoparticle structural disorder in plasmon-polariton formation and the strength of light-matter coupling in plasmonic assemblies.  Understanding how disorder affects optical properties has practical implications since methods for assembling low-defect nanoparticle superlattices are slow and scale poorly. Modeling realistic disorder requires large system sizes, which is challenging using conventional electromagnetic simulations. We employ Brownian dynamics simulations to construct large-scale nanoparticle multilayers with controlled structural order. We investigate their far- and near-field optical response using a superposition T-matrix method with two-dimensional periodic boundary conditions. We find that while structural disorder broadens the polaritonic stop band and the near-field hot-spot distribution, the polariton dispersion and coupling strength remain unaltered. To understand the effects of nanoparticle composition, we consider assemblies with Drude model particles mimicking gold or tin-doped indium oxide (ITO) nanocrystals. Losses due to higher damping in ITO nanocrystals prevent their assemblies from achieving the deep, strong coupling of gold nanoparticle multilayers, although the former still exhibit ultrastrong coupling. Finally, we demonstrate that while computationally efficient mutual polarization method calculations employing the quasistatic approximation modestly overestimate the strength of collective plasmon coupling in these assemblies, they reproduce the polariton dispersion relations determined by electrodynamic simulations.
\end{abstract}

\begin{tocentry}
\includegraphics[width=\textwidth]{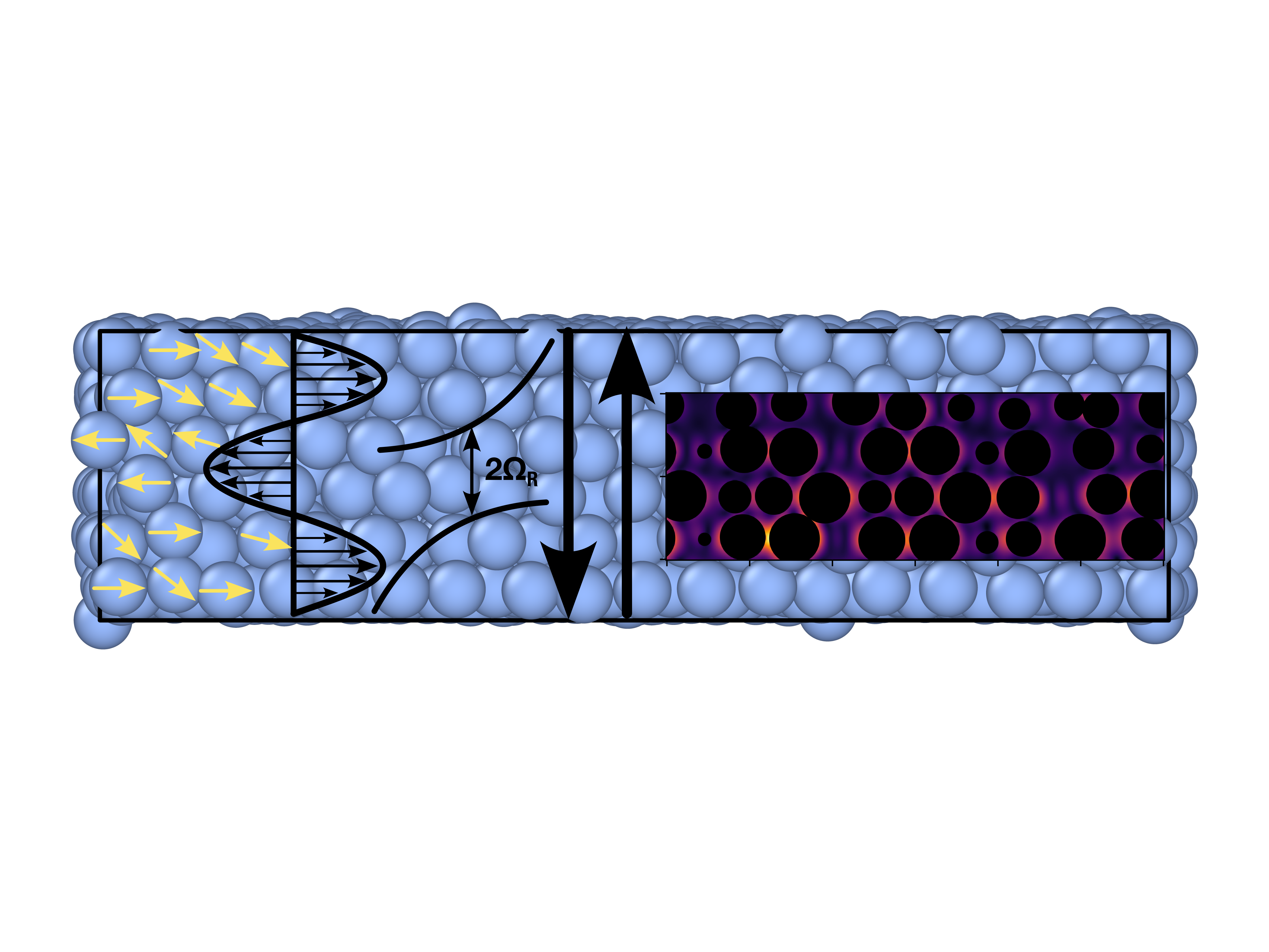}
\end{tocentry}


\section*{Introduction}

\begin{figure}[h]
    \centering
    \includegraphics[width=\textwidth]{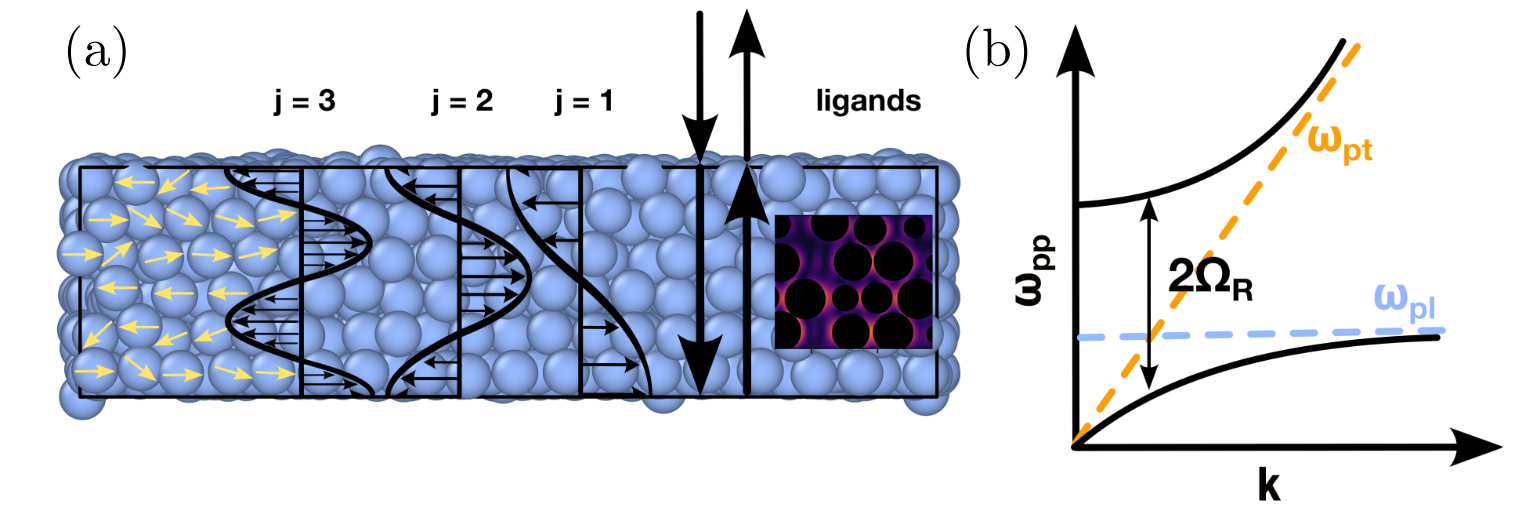}
    \caption{(a) Schematic of plasmon polaritons in a three-dimensional disordered nanoparticle assembly. Nanoparticle dipole moments are indicated by yellow arrows. Standing waves are labeled by their index $j$. Inset: Near field intensity map at the $j = 1$ polariton mode. 
    (b) Schematic diagram of plasmon polariton dispersion. The orange dashed line indicates photon dispersion, and the blue dashed line indicates the collective plasmon dispersion. Strong light-matter coupling is shown by the Rabi splitting ($2\Omega_{\rm R}$) between the upper and lower polariton branches arising from an anti-crossing of the photon and plasmon dispersions.}
    \label{fig:Fig1}
\end{figure}

Strong light-matter interactions can enhance the performance of materials in solar energy and photoelectric conversion devices\cite{shi_water}, perfect absorbers and photovoltaics \cite{hagglund_photo,chang2023wavelength,chang2025ultrastrong}, molecular sensors, \cite{ameling2013microcavity} and many other applications. 
One example is the assembly of three-dimensional superlattices (SLs) of gold (Au) nanoparticles (NPs) for surface-enhanced Raman scattering or surface-enhanced infrared absorption \cite{mueller2021surface}.
Fabry-P{\'e}rot cavity modes due to confinement of electromagnetic fields within the SL can couple to the assembly's collective electron oscillations, producing hybrid plasmon-polariton states.\cite{Huang_PRL,Mueller_2020,mueller2021surface,Park_PNAS}. 
A schematic diagram is presented in Figure \ref{fig:Fig1}(a), where the standing waves of the first three cavity modes, indexed by $j$, are shown. 
The photon field induces dipolar oscillations in NPs (schematically indicated in Figure~\ref{fig:Fig1}(a) by yellow arrows), which interact with each other via near-field coupling, leading to the formation of a collective plasmon resonance.
A strong polaritonic coupling between this collective plasmon and the cavity mode is evident from the splitting between the upper and lower polariton branches of the dispersion relation, referred to as Rabi splitting \cite{garcia_ultrastrong, khitrova_rabi}, along with the emergence of a polaritonic stop band between the branches where no modes are present \cite{Huang_PRL,Park_PNAS,Mueller_2020}, as schematically illustrated in Figure \ref{fig:Fig1}(b). 

Plasmonic NPs are attractive building blocks for photonic metamaterials due to their strong and highly tunable interaction with light, which originates from collective excitations of free electrons, known as localized surface plasmon resonance\cite{Agrawal2018}. 
When assembled into a densely packed structure, the coupling of neighboring NPs' surface plasmons leads to the formation of a lower energy collective (transverse) plasmon resonance (CPR) with enhanced near-field intensities in the small gaps between NPs \cite{yang2010ability,ye2013tunable,solis2014toward,Ross_2015,Park_PNAS,ross2015nanoscale,garcia2019plasmonic,gm2019plasmonic,wang2022surface,Berry_Milliron_2024,Sherman2023, Sherman_PlasmonRuler_2023,green_2024}. 
For surface-enhanced spectroscopies, the number of layers in an NP thin film can be adjusted to tune or detune relative to vibrational modes of interest, engineering environments for sensitive molecular detection \cite{Mueller_2021,Arul_2022,Chang_SEIRA_2024,chang2025ultrastrong}. 
Because noble metals exhibit localized surface plasmonic responses in the visible range, they are not ideally suited for infrared (IR) sensing or thermal management applications. 
In contrast, plasmonic doped metal oxide NPs are transparent in the visible range and offer broadly tunable resonance frequencies in the near-IR (NIR) or mid-IR regions. 
Tin-doped indium oxide (ITO) nanocrystals, in particular, are excellent NP building blocks for IR-sensitive metamaterials \cite{Agrawal2017,staller2019, Jansons2016, Kanehara2009,xi2018localized,Matsui2022,Sherman2023,kang2022colorimetric,kang2023modular,kang2023structural,chang2023wavelength, Kihoon_Meta_2023,chang2024plasmonic,chang2025ultrastrong}. 
Their resonance frequency can be tuned precisely by synthetically adjusting their dopant composition, enabling the optimization of light-matter interactions to meet a specific design requirement. 
However, quantitative design of optical properties also requires an understanding of how both particle characteristics (e.g., size and shape uniformity) and assembly structure (e.g., particle spacing and coordination, defectivity, etc.) influence polaritonic coupling \cite{Mueller_2018,Schulz_2020_NatComm,Schulz_Review_2021}.

Plasmon polaritons are sensitive to the particle sizes and interparticle spacings in an NP assembly.\cite{Mueller_2018} Recent studies of Au NP SLs have demonstrated that with robust synthesis of uniform particles and assembly protocols producing well-ordered SLs, plasmon-polariton modes can be clearly identified and characterized in local regions with well-defined layer thickness.\cite{Mueller_2018, Mueller_2020, mueller2021surface} 
However, synthesizing highly uniform particles is challenging, and assemblies of particles with sufficiently broad or irregular particle shape and size distributions do not support plasmon polaritons.\cite{Schulz_2020_NatComm}.
Analogously, sharp diffractive peaks, known as surface lattice resonances, in lithographically fabricated perfectly ordered Au NP arrays become progressively broadened and eventually vanish when pseudo-random positional disorder is introduced\cite{Auguie_2009}. Since particles with nonuniform sizes or shapes naturally assemble into more disordered SLs, it is challenging to determine which aspect is more critical for forming plasmon polaritons:  
synthesizing uniform NPs or carefully assembling them into highly ordered structures. An experimental study of Au NP multilayer assemblies featuring stacked disordered layers on a Au mirror showed evidence of plasmon polaritons, suggesting that assembling perfectly ordered NP SLs is not a strict requirement.\cite{Arul_2022} More recently, both experiments and simulations of ITO nanocrystal multilayers in an open optical cavity architecture produced plasmon polaritons, despite the layers exhibiting significant structural disorder.\cite{chang2025ultrastrong} These studies raise fundamental and practical questions about the role disorder plays in the formation of plasmon-polaritons in metallic NP SLs. Here, we aim to systematically investigate how structural disorder in NP assemblies influences plasmon-polariton formation and polaritonic coupling strength using computer simulation.

Simulation enables systematic exploration of experimentally relevant systems with precise control over individual particle characteristics, providing a pathway to a comprehensive understanding of how structural order influences plasmon polaritons in NP assemblies. 
Recently, we investigated the plasmonic response of structurally disordered NP monolayers \cite{green_2024} using the mutual polarization method (MPM) \cite{Sherman2023,Sherman_PlasmonRuler_2023,Kihoon_Meta_2023}, a moment-based approach for determining the polarization and other optical properties of spherical NP assemblies in the quasistatic limit. MPM, a generalized coupled-dipole method,\cite{dey2024plasmonic,markel2019extinction,herkert2023influence} determines how incident light and the spatial arrangement of the particles with known optical properties impact mutual polarization of the particle dipoles (and quadrupoles) due to collective oscillation of their conduction electrons.
The ability of MPM computations to converge quickly, even for large system sizes with periodic boundary conditions and arbitrary particle configurations (including particle overlaps), makes them well-suited for analyzing data from particle-based (e.g., Brownian dynamics) simulations and addressing questions about the implications of structural disorder for optical properties.\cite{Sherman2023,sherman2024review}

While the quasistatic approximation is appropriate for analyzing NP monolayers, its predictions may be expected to become increasingly inaccurate for multilayer structures with dimensions approaching the wavelength of light. To accurately study plasmon polaritons in multilayered NP arrangements with disorder, a simulation method must accommodate both large numbers of particles to properly capture structural heterogeneity while accounting for position-dependent phase variations in the electric field. To account for these effects, we compute electromagnetic properties using SMUTHI (Scattering by
Multiple Particles in Thin-film Systems), a python-based software package capable of leveraging superposition T-matrix (transition-matrix) calculations with Ewald sums to analyze the propagation of light through three-dimensional layered assemblies with finite thickness in the $z$ direction and periodic boundary conditions in the transverse $xy$ plane.\cite{egel2017celes,egel2021smuthi, theobald2021simulation}. 
SMUTHI computations are considerably more expensive than those of MPM, but they are sufficiently fast to enable the electrodynamic analysis of NP assemblies with hundreds to thousands of plasmonic particles, which is sufficient to account for the disorder present in many experimental multilayers.\cite{chang2025ultrastrong}
To understand the role of electrodynamic effects on plasmon-polaritons in multilayer NP structures, we compare SMUTHI predictions for select cases to the quasistatic predictions of MPM. 

To study the implications of structural disorder for optical properties, we introduce a simple, two-step method for generating NP configurations for multilayer assemblies. In the first step, dense monolayers of monodisperse spherical NPs are formed by a two-dimensional compression of a dilute monolayer. The degree of structural order in the layers is controlled by the compression rate, with slower rates leading to more ordered layers. Then the layers are equilibrated into a stacked assembly. This process, loosely analogous to liquid-air, layer-by-layer SL assembly protocols performed experimentally,\cite{chang2025ultrastrong} is used to generate a series of three-dimensional films exhibiting varying degrees of structural disorder. We compute the optical properties of these films, as well as SLs with perfect ordering, assuming the optical properties of isolated NPs can be described by the Drude model. We consider Drude parameters that mimic NPs of two materials:  Au, a noble metal which has been extensively studied in the context of NP SLs with plasmon polaritons,\cite{Mueller_2020,mueller2021surface,Schulz_2020_NatComm,Arul_2022} and ITO, a plasmonic transparent conductive oxide material of interest because of its synthetic tunability and strong interaction with NIR and mid-IR light.\cite{Agrawal2017,staller2019, Jansons2016, Kanehara2009,xi2018localized,Matsui2022,Sherman2023,kang2023modular,kang2023structural,chang2023wavelength, Kihoon_Meta_2023,chang2024plasmonic,chang2025ultrastrong}       

Our simulations of the optical properties of model Au and ITO NP assemblies using SMUTHI indicate that, though structural disorder contributes to modest broadening and weakening of the polaritonic stop band and the disappearance of a few plasmon-polariton modes in the stop band's vicinity, disorder does not change the polariton dispersion relation relative to that of a perfectly ordered SL. In other words, plasmon-polariton dispersion in plasmonic NP SLs has high defect tolerance. In terms of nanoparticle composition, our modeling indicates that ITO nanocrystal SLs are expected to exhibit weaker plasmon-polariton coupling compared to the deep, strong coupling of Au NP SLs, due to the higher damping of individual ITO nanocrystals; however, they still exhibit ultrastrong coupling.  This difference is counterbalanced by the increased flexibility of ITO nanocrystals, which have resonances that can be sensitively tuned via dopant concentration independent of NP size. Finally, we find that the plasmon-polariton dispersion curves predicted by MPM in the quasistatic approximation closely match those computed from SMUTHI, which accounts for electrodynamic effects. This close correspondence suggests that quasistatic computational approaches, which are simple and have low computational cost, can be helpful in the design of SLs with plasmon polaritons, e.g., as a screening tool to identify promising parameters for full-wave electrodynamic calculations.

\section*{Model and Methods}

\subsection*{Generation of perfect SLs}

Perfectly ordered structures were generated by placing particles on (111) planes of an FCC lattice. 
The separation between two layers is $d_{111}=L/\sqrt{3}$, where the lattice constant of the FCC lattice is given by $L=\sqrt{2}(d+l_g)$, $d$ being the diameter associated with a NP's plasmonic core and $l_g$ being the length of the gap between two nearest NP cores. 
We considered $d=30$~nm and $l_g=3$~nm for the Au SL. 
For ITO SL, these values were chosen to be $d=90$~nm and $l_g=9$~nm, respectively. The larger scaling was selected for the ITO SL to facilitate investigating a similar regime of cavity lengths relative to the plasmon resonance wavelength while maintaining the number of NP layers, to avoid adding unnecessary computational expense.
The $d/(d+l_g)$ ratio is held fixed in both cases so that the area fraction for both SLs is the same and is $\phi_0 \simeq 0.75$. 

\subsection*{Generation of assemblies with disorder}

We prepared multilayer NP assemblies with disorder via Brownian dynamics simulations using HOOMD-blue v4.8.2 \cite{Anderson2020hoomd}. 
The interaction between hard-sphere particles was implemented as \cite{Heyes1993}
\begin{equation}
  U_{\rm HS}(r) =  H(\sigma-r)\dfrac{\gamma_h}{4\Delta t}(r-\sigma)^2
\end{equation}
where $H$ is the Heaviside step function, $\sigma=2a_0$ is the nearest approach between two particles, i.e., $a_0$ is their radius, $\gamma_h$ is the Stokes-Einstein hydrodynamic drag coefficient, and $\Delta t$ is the time step. In an experimental system, NPs are coated with a ligand shell, so their `thermodynamic' diameter $\sigma$ is larger than the diameter of their plasmonic core, $d$.
We have considered $\Delta t=10^{-4}\ \tau_{D}$, where $\tau_{D}=\gamma_h a_0^2/k_B T$ is the diffusion time of an NP. As described elsewhere,\cite{Sherman2023} this implementation of the Heyes-Melrose algorithm\cite{Heyes1993} generates hard-sphere configurations with minimal overlaps in Brownian dynamics even at high density. 

\begin{figure}[h]
    \centering
    \includegraphics[width=\textwidth]{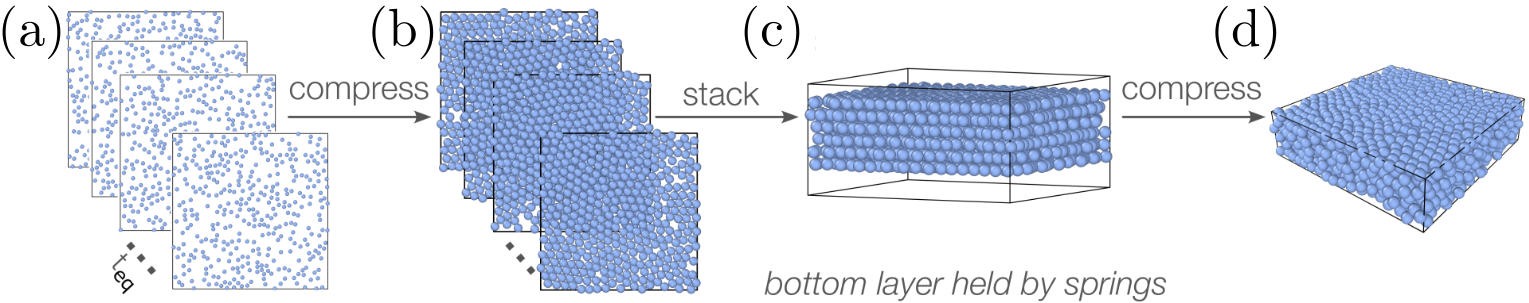}
    \caption{Protocol for generating disordered multilayers. (a) Low area fraction ($\phi = 0.20$) hard-sphere monolayer configurations extracted from Brownian dynamics simulations at different time points $t_{\rm eq}$.
    (b) Monolayers compressed to an area fraction $\phi \simeq 0.90$ over a time $t_{2d}$, which determines their structural order. Slow compressions lead to more ordered monolayers. (c) Monolayers are stacked, with bottom-layer particles tethered to their initial position by springs, and (d) compressed into a final 3D film configuration. Snapshots are visualized using Ovito \cite{ovito}.}
    \label{fig:Fig2}
\end{figure}

Initially, monolayers were generated from equilibrated particle configurations at a low area fraction of $\phi=0.2$ [Figure \ref{fig:Fig2}(a)] with two-dimensional (2D) periodic boundaries. The monolayers were then compressed under Brownian dynamics to a final high area fraction of $\phi \simeq 0.9$ [Figure \ref{fig:Fig2}(b)]. By choosing different values of the compression time $t_{2d}$, monolayers with varying degrees of structural order were obtained (Table S1).\cite{green_2024} As described below, structural order was quantified using the local hexatic bond-orientational order parameter, $\psi_6$. Fast compressions generated monolayers with less structural order,\cite{lubachevsky1991disks,truskett2000} characterized here by a lower $\psi_6$. Multiple monolayers compressed from distinct low-density configurations for the same $t_{2d}$ were stacked on top of one another in a 3D simulation cell [Figure \ref{fig:Fig2}(c)], which was then vertically compressed for time $t_{3d}$ to produce the final 3D multilayer assembly [Figure \ref{fig:Fig2}(d) and Table S1]. During this step, the vertical displacement of particles in the bottommost layer was restrained by tethering them to their initial positions with a harmonic spring potential
\begin{equation}
    U_{\rm s}(r) = \dfrac{1}{2} kr^2,
\end{equation}
where $r^2$ is a particle's square displacement and $k=2000$ $k_BT/a_0^2$ is the force constant. Finally, a constant downward force of magnitude $9.8$ $k_BT/a_0$ was applied to all particles to further restrict their motion in the vertical direction.

Optical simulations were carried out on disordered NP films prepared as described above with six layers and $361$ particles per layer. The occupied area fractions of each layer ($\phi \simeq 0.9$) reflect the thermodynamic diameter $\sigma$ of the NPs, which defines their interaction in Brownian dynamics. However, for optical simulations, we consider only the inner plasmonic particle diameter $d$, which corresponds to a lower area fraction of $\phi_0 \simeq 0.75$. 
Specifically, an NP core with $d = 30$~nm surrounded by a ligand shell of thickness $l_g/2 = 1.5$~nm, or a NP with a core of $d = 90$~nm and a $l_g/2 = 4.5$~nm shell, both have a core area fraction of $\phi_0 \simeq 0.75$ in the optical simulations, matching those of perfect SLs described above. 

\subsection*{Structure characterization}

To characterize 2D and 3D structures, the average hexatic ($\psi_6$) and Steinhardt six-fold ($q_6$) local bond-orientational order parameters,\cite{lotito2020pattern} respectively, were calculated using freud (Table S1)\cite{freud2020}. 
To calculate $\psi_6$ within the planar NP layers of the 3D configurations, the $z$ dimension was binned into sections (i.e., six for a six-layered structure), and a two-dimensional Voronoi construction was carried out for each layer to determine each particle's neighbors within its bin.\cite{lotito2020pattern}
The local hexatic bond-orientational order parameter for a particle $i$ was calculated as
\begin{equation}
    \psi_{6,i} = \dfrac{1}{N_{b,i}}\sum_{k=1}^{N_{b,i}} e^{6j\alpha_{ik}},
\end{equation}
where $N_{b,i}$ is the particle's number of neighbors, and $\alpha_{ik}$ is the angle between the vector separating particles $i$ and $k$ and an arbitrary reference axis. 
The average local hexatic order parameter for a layer is computed by averaging $|\psi_{6,i}|$ over each particle $i$ in the layer.
\begin{equation}
    \psi_6 = \langle |\psi_{6,i}| \rangle
\end{equation}

The Steinhardt six-fold bond-orientational order parameter for particle $i$ is given by,\cite{steinhardt1983bond}
\begin{equation}
    q_{6}(i) = \sqrt{\dfrac{4\pi}{2(6)+1}\sum_{m=-6}^{6}|q_{6m}(i)|^2}
\end{equation}
with
\begin{equation}
    q_{6m}(i) = \dfrac{1}{N_{b,i}^{(q)}}\sum_{k=1}^{N_{b,i}^{(q)}} Y_{6m}(\Theta(\vec{r}_{ik}), \Phi(\vec{r}_{ik}))
\end{equation}
where $Y_{lm}$ are spherical harmonics, $\vec{r}_{ik}$ is the separation vector between particles $i$ and $k$, and $\Theta(\vec{r}_{ik})$ and $\Phi(\vec{r}_{ik})$ are the corresponding spherical angular coordinates. 
Here, $N_{b,i}^{(q)}$ is the number of neighbors of the $i$th particle considered for this 3D order parameter. This number considers all particles (in any layer) that meet the criterion $r_{ik} \le 1.05\sigma$. 
For a given configuration, the reported $q_6$ was averaged over all the particles. For a perfect FCC lattice, $q_6$ is $q_{6,{\rm FCC}} \simeq 0.58$.\cite{steinhardt1983bond,rein1996numerical}

The parallel (in-plane) radial distribution function $g_{\parallel}(r)$ was computed using conditional binning, as previously reported \cite{Singh_2022}. 
Briefly, $g_{\parallel}$ is computed for particle pairs whose vertical interparticle distance $r_{\perp} \leq d/2$, and $g_{\parallel}$ is defined as:
\begin{equation}
    g_{\parallel}(r) = \dfrac{2}{\rho N \pi d^2 h} \sum_{ij}^{N}\begin{cases}
    1,& \text{if } r-\frac{h}{2}\leq r_{ij} < r+\frac{h}{2} \text{ and } r_{\perp} \leq d/2\\
    0,              & \text{otherwise}
    \end{cases}
\end{equation}
with $\rho$ as the particle number density and $h = 0.004d$ the bin spacing. 

\subsection*{Optical simulation}

For optical simulations, we consider normal incidence of s-polarized light; i.e., the propagation wavevector $\vec{k}$ is perpendicular to the NP layers and parallel to the (111) lattice direction.  We computed the optical properties of NC assemblies using SMUTHI. This Python-based package utilizes superposition transition matrix (T-matrix) and Ewald sums to solve Maxwell equations in the frequency domain for 3D layered systems with interfaces \cite{egel2021smuthi}, applying 2D periodic boundary conditions in directions parallel to the NP layers \cite{theobald2021simulation}. 
SMUTHI can consider higher-order multipoles, but at an increasingly expensive computational effort, especially for the large systems required to mimic realistic structural disorder. 
Here, we primarily report calculations that only consider dipole excitation of plasmonic NPs, selecting the maximum multipole order $l=1$ in most cases. 
Contributions from higher-order electrical modes in individual plasmonic nanoparticles of small size ($< 100$~nm) are expected to be small\cite{Maierbook, barros2021}. 
Based on the far-field optical properties, near-field enhancement can also be computed. Since periodic boundaries are not supported for near-field calculations in SMUTHI, we focused on the interior region far from the $x$ and $y$ simulation cell boundaries, where edge effects were absent. 
Field intensities were computed at uniformly spaced points with a spatial resolution of $1$~nm over the selected plane section. 
All calculations used a three-medium setup in SMUTHI where the top and bottom media have the same permittivity, $\varepsilon_m$, as the interstitial `ligand' media surrounding the NPs.  
The frequency-dependent permittivity of plasmonic NP cores was modeled using a Drude dielectric function
\begin{equation}
    \varepsilon_p(\omega) = \varepsilon_{\infty} - \dfrac{\omega_{\rm p}^2}{\omega^2+i\gamma\omega}, \label{eq:Drude}
\end{equation}
where $\varepsilon_{\infty}$ is the high frequency dielectric constant, $\omega_{\rm p}$ is the NP plasma frequency, $\omega$ is the angular frequency, and $\gamma$ is the damping rate. 
Here, permittivities are expressed considering the free-space permittivity $\varepsilon_0=1$. 
The chosen model parameters are provided in Table \ref{table:system_parameters}. 
\begin{table}[h!]
\centering
 \begin{tabular}{|c c c|} 
 \hline
 Quantity & Au SL & ITO SL \\
 \hline
 Plasma frequency ($\omega_{\rm p}$) & \multirow{2}{*}{68558} & \multirow{2}{*}{15000} \\
 (cm\textsuperscript{-1})            &                        &                        \\ 
 \hline
 Damping ($\gamma$)                  & \multirow{2}{*}{387}   & \multirow{2}{*}{800}   \\
 (cm\textsuperscript{-1})            &                        &                        \\ 
 \hline
 High-frequency dielectric           & \multirow{2}{*}{1}     & \multirow{2}{*}{4}     \\
 constant ($\varepsilon_{\infty}/\varepsilon_0$)              &                        &  \\ 
 \hline
 Dielectric constant of the          & \multirow{2}{*}{2.00}  & \multirow{2}{*}{2.00}  \\
 surrounding medium ($\varepsilon_{m}/\varepsilon_0$)         &                        &  \\
 \hline
 NP plasmonic core diameter ($d$)                   & \multirow{2}{*}{30}    & \multirow{2}{*}{90}    \\
 (nm)                                &                        &                        \\
 \hline
 Gap between nearest NPs             & \multirow{2}{*}{3}     & \multirow{2}{*}{9}     \\
 in perfect SL ($l_g$) (nm)          &                        &                        \\
 \hline
 Plasmonic core area fraction of            & \multirow{2}{*}{0.75}  & \multirow{2}{*}{0.75}  \\
 a single layer ($\phi_0$)           &                        &                        \\
 \hline
 \multirow{2}{*}{Angle of incidence ($\theta$)} & \multirow{2}{*}{0\degree} & \multirow{2}{*}{0\degree} \\
                                     &                        &                        \\
 \hline
 \end{tabular}
 \caption{Model parameters used for optical simulations.} \label{table:system_parameters}
\end{table}

The relevant far-field quantities include reflectance ($R$), transmittance ($T$), and absorptance ($A$). 
These are defined as follows:
\begin{equation}
    R = \dfrac{\left| \vec{E}_r \right|^2}{\left| \vec{E}_0 \right|^2}, \quad T = \dfrac{\left| \vec{E}_t \right|^2}{\left| \vec{E}_0 \right|^2}, \quad \text{and } A = 1 - T - R, \label{eq:RTA}
\end{equation}
where $\vec{E}_0$, $\vec{E}_r$, and $\vec{E}_t$ are the incident, reflected, and transmitted electric fields, respectively.

\subsection*{Theoretical model for plasmon polaritons} 

The hybridized states of plasmon polaritons, which originate from strong light-matter interaction in plasmonic NP assemblies, can be described by a quantum mechanical model that accounts for NP dipole moment excitations, photon field retardation, and wavevector-dependent polarization of light \cite{barros2021, Mueller_2020}. 
Within this formalism, a Hopfield-type interaction \cite{hopfield1958} between the photon field, characterized by the cavity-mode frequency $\omega_{\rm cav}(\vec{k})$, and the transverse collective plasmon, characterized by the CPR frequency $\omega_{\rm CPR}$, yields a dispersion relation for the hybridized plasmon polariton states, $\omega_{\rm pp}(\vec{k})$,\cite{barros2021, Mueller_2020}
\begin{equation}
     \omega_{{\rm  pp}}^{4}(\vec{k}) - \omega_{{\rm  pp}}^{2}(\vec{k}) \left[ \omega_{{\rm  cav}}^{2}(\vec{k}) + \omega_{{\rm  CPR}}^{2} + 4 \Omega_{{\rm R}}^{2} \right] + \omega_{{\rm  cav}}^{2}(\vec{k}) \omega_{{\rm  CPR}}^{2} = 0 \nonumber \\
    , \label{eq:HF_disp1}
\end{equation}
which can be recast as 
\begin{equation}
     \omega_{\rm pp}\left( \vec{k} \right) = \sqrt{ \dfrac{1}{2} \left[ \left( \omega_{\rm  cav}^{2} + 4 \Omega_{\rm R}^2 + \omega_{\rm CPR}^2  \right) \pm \sqrt{ \left( \omega_{\rm  cav}^{2} + 4 \Omega_{\rm R}^2 + \omega_{\rm CPR}^2  \right)^2 - 4 \omega_{\rm  cav}^{2} \omega_{\rm CPR}^2 } \right] }. \label{eq:HF_disp}
\end{equation}
Here, the Fabry-P{\'e}rot cavity modes are described by the photon dispersion,
\begin{equation}
    \omega_{\rm cav}(\vec{k}) = ck/n_{\rm eff}. \label{eq:photon_dispersion}
\end{equation}
The wavenumber of incident light $k=|\vec{k}|=2\pi/\lambda$ depends on its wavelength $\lambda$, $c$ is the speed of light in vacuum, and $n_{\rm eff}$ is the refractive index of the superlattice in the absence of the metallic resonance. 
Note that $n_{\rm eff}$ has contributions from both the surrounding medium and the screening effect by the bound charges in the NPs, so its value depends on the NP composition (i.e., Au or ITO in this study). 
We approximate $n_{\rm eff}$ as the mean refractive index for the photon field when the plasmonic character of NPs is absent, and express $n_{\rm eff}$ as \cite{Mueller_2020}
\begin{equation}
    n_{{\rm eff}} = (1-F)\sqrt{\varepsilon_{m}}+F\sqrt{\varepsilon_{d}},
    \label{eq:neff}
\end{equation}
where the volume fraction of the NPs is
\begin{equation}
    F = \left( \dfrac{d}{d+l_g} \right)^3 F_0, \label{eq:F}
\end{equation}
Here, $F_0=\pi/(3\sqrt2)$ is the volume fraction for zero interparticle gap, $\varepsilon_{m}$ is the permittivity of the surrounding medium, and $\varepsilon_d$ is the background dielectric constant due to the bound charges of the NPs. 

The strength of the light-matter interaction is related to the so-called Rabi \emph{splitting} $2\Omega_{\rm R}$, the minimum frequency difference between upper ($\omega_{\rm pp}^{+}$) and lower ($\omega_{\rm pp}^{-}$) polariton branches of the dispersion relation over all wave vectors $\vec{k}$ in the first Brillouin zone along a fixed direction $\hat{k}$ from the $\Gamma$ point \cite{Lamowski2018}. In the perfectly ordered SLs studied here, we consider the $\Gamma L$ direction, which corresponds to the (111) direction of the FCC lattice. Within the Hopfield Hamiltonian model dispersion (eq~\ref{eq:HF_disp}), it can be shown that the Rabi splitting is equal to the frequency difference between the polariton branches where cavity and CPR modes coincide, i.e., $\omega_{\rm cav}(\vec{k}) = \omega_{\rm CPR}$. 

To characterize the light-matter coupling of Au and ITO NP assemblies, we fit eq~\ref{eq:HF_disp} to the plasmon-polariton dispersion relations computed by SMUTHI, considering $n_{\rm eff}$, $\omega_{\rm CPR}$, and $\Omega_{\rm R}$ as free parameters. 
The ratio of the Rabi \emph{frequency} $\Omega_{\rm R}$ to the CPR frequency $\omega_{\rm CPR}$ is a reduced coupling strength, 
\begin{equation}
    \eta = \dfrac{\Omega_{\rm R}}{\omega_{\rm CPR}}, \label{eq:eta}
\end{equation}
which serves as a useful parameter for comparing different materials and systems.

The dipole plasmon frequency of a lossless ($\gamma=0$) Drude NP can be written as
\begin{equation}
    \omega_{{\rm NP}} = \dfrac{\omega_{{\rm p}}}{\sqrt{\varepsilon_{\infty}+2\varepsilon_{m}}}, \label{eq:omega_NP_Mie}
\end{equation}
which occurs at the resonance condition $\varepsilon_{p}(\omega_{\rm NP}) = -2\varepsilon_m$ for dipole plasmons.\cite{bohren2008absorption} 
In a dense collection of these NPs, the hybridized state of the collective plasmon can be represented as \cite{Mueller_2020, barros2021, Lamowski2018}
\begin{equation}
    \omega_{\rm CPR}(\vec{k}) = \omega_{\rm NP} \sqrt{1 + g f(k)}. \label{eq:omega_CPR_micro}
\end{equation}
Here, $g$ is a coupling factor that depends on the NP volume fraction $F$ [see eq~\ref{eq:F}], the effect of background screening by bound charges in NPs, and the surrounding medium. The coupling factor is given by
\begin{equation}
    g = \dfrac{3}{4} \left(\dfrac{3\varepsilon_m}{\varepsilon_{d}+2\varepsilon_m}\right) F. 
    \label{eq:g}
\end{equation}
The dispersion of the collective plasmon is dictated by the factor $f(k)$ that depends on the structure of the NP assembly and the plasmon polarization, which is the transverse plasmon in our case. 
The quantity $f(k)$ can be calculated from the lattice summation of the plasmonic dipoles \cite{Lamowski2018}. 
The Rabi frequency $\Omega_{\rm R}$, quantifying the coupling between the transverse collective dipole plasmon and the cavity mode, is related to the coupling factor $g$ by \cite{Mueller_2020, barros2021}
\begin{equation}
    \Omega_{\rm R} = \omega_{\rm NP} \sqrt{g}. \label{eq:Omega_R_micro}
\end{equation}
Thus, an analytical expression for the reduced coupling constant $\eta$ can be obtained by combining eqs \ref{eq:eta}, \ref{eq:omega_CPR_micro}, and \ref{eq:Omega_R_micro} as
\begin{equation}
    \eta = \left[ \dfrac{4}{3F} \dfrac{\varepsilon_{d} + 2\varepsilon_m} {3\varepsilon_m} + f(k) \right]^{-1/2}. 
    \label{eq:eta_Th}
\end{equation}

As demonstrated in earlier studies \cite{Lamowski2018, Mueller_2020}, $f(k)$ may be approximated by a constant over the wavevector range in the first Brillouin zone along the $\Gamma L$ direction, the direction relevant for our study. 
In our study, we evaluated $f(k)$ from eq~\ref{eq:omega_CPR_micro}, by estimating $\omega_{\rm NP}$ from eq~\ref{eq:omega_NP_Mie} and $\omega_{\rm CPR}$ from the extinction peak position of an assembly of NPs in a perfect monolayer (see Figure~S9). 
With this value of $f(k)$, we estimate $\eta$ from eq~\ref{eq:eta_Th} and compare it with the estimations obtained from the fitting of the dispersion data. 

The dispersion relation for polaritonic assembly of metallic NPs can be expressed in terms of an effective permittivity $\varepsilon_{\rm eff, pp}$,\cite{Mueller_2020, Park_PNAS} 
\begin{equation}
    \omega_{\rm pp} = ck/\sqrt{\varepsilon_{\rm eff, pp}}. \label{eq:eff_pp_dispersion}
\end{equation} 
The polariton dispersion relation in eq~\ref{eq:HF_disp} can be obtained from eq~\ref{eq:eff_pp_dispersion} by expressing $\varepsilon_{\rm eff, pp}$ as\cite{Mueller_2020, Mills1974}
\begin{equation}
    \varepsilon_{\rm eff, pp} (\omega_{\rm pp}) = \varepsilon_{\infty, \rm pp} + G \dfrac{\omega_{\rm CPR}^2}{\omega_{\rm CPR}^2 - \omega_{\rm pp}^2}, \label{eq:eps_eff_pp}
\end{equation}
and identifying $\varepsilon_{\infty, \rm pp}=n_{\rm eff}^2$. 
Here, the oscillator strength $G=4 n_{\rm eff}^2 \tilde{\eta}^2$, where $\tilde{\eta}$ represents the reduced coupling strength for an ideal material with zero intrinsic loss ($\gamma=0$). 
For a real system with $\gamma \ne 0$, it may be estimated by replacing $\tilde{\eta}$ by $\eta$, i.e., using the relation $G=4 n_{\rm eff}^2 \eta^2$, which can help compare metamaterials based on their plasmon-polaritonic coupling strength. 
Note that, in eq \ref{eq:eps_eff_pp}, $\sqrt{G}$ is the value of the plasma frequency of the oscillator, $\omega_{\rm p}$, in units of $\omega_{\rm CPR}$. 
This connection of $G$ with $\omega_{\rm p}$ leads to the relation of the reduced coupling strength 
\begin{equation}
    \tilde{\eta} = \dfrac{1}{2n_{\rm eff}} \left( \dfrac{\omega_{\rm p}}{\omega_{\rm CPR}} \right), \label{eq:eta_tilde}
\end{equation}
and characterizes the plasmon polaritonic coupling strength of a metamaterial in terms of quantities that are either known or can be estimated theoretically without the need for detailed electromagnetic calculations or experiments. 
Interestingly, the dimensionless quantity $\tilde{\eta}$ encapsulates the key aspects of the system, viz., the individual NP property via $\omega_p$, the collective plasmon property via $\omega_{\rm CPR}$, and the cavity characteristics via $n_{\rm eff}$. 
Thus, estimating $\tilde{\eta}$ provides a practical way to gauge the expected coupling strength of a metamaterial design before experimental investigation. 
However, it should be noted that $\tilde{\eta}$ can overestimate the actual reduced coupling strength, $\eta$, in a real system with non-negligible intrinsic loss ($\gamma \ne 0$). 

Polariton states (i.e., modes) can be identified from far-field optical properties as minima in the reflectance spectra. The polariton frequency $\omega_{{\rm pp}, j}$ corresponding to the $j$th mode is identified by the $j$th lowest frequency minima ($j=1, 2, 3, \dots$) below the stop band (lower polariton branch) or above the stop band (upper polariton branch), and the correspnding dispersion relations (eq~\ref{eq:eff_pp_dispersion})  are given by $\omega_{{\rm pp}, j} = ck_j/\sqrt{\varepsilon_{\rm eff, pp}}$. The polariton wavenumber $k_j$ corresponding to the $j$th mode is defined as $k_j = 2\pi / \lambda_{{\rm pp}, j}$, with the polariton wavelength $\lambda_{{\rm pp}, j}$ satisfying the resonance condition
\begin{equation}
    h = j\left( \dfrac{\lambda_{{\rm pp}, j}}{2} \right), \label{eq:pp_wl_resonance}
\end{equation}
where $h$ is the film thickness. 

\section*{Results}

Control over NP SL thickness (i.e, cavity height $h$) is crucial for establishing and characterizing polariton modes and can be achieved experimentally through evaporative liquid-air interface assembly approaches.\cite{dong2010binary,schulz2017size,Mueller_2020,Schulz_2020_NatComm,Mueller_2021} A simplified computational version of this multilayer assembly process is described in detail in the Methods section. 
The 3D assemblies generated for this work intentionally span a range of structural disorder (i.e., defectivity) so that the effects of disorder on optical properties can be studied. 
To quantify defectivity, we computed a local 2D (intralayer) bond-orientational order parameter $\psi_6$ and radial distribution function $g_{\parallel} (r)$ and a 3D bond-orientational order parameter $q_6$.

\begin{figure}[h!]
    \centering
    \includegraphics[width=\textwidth]{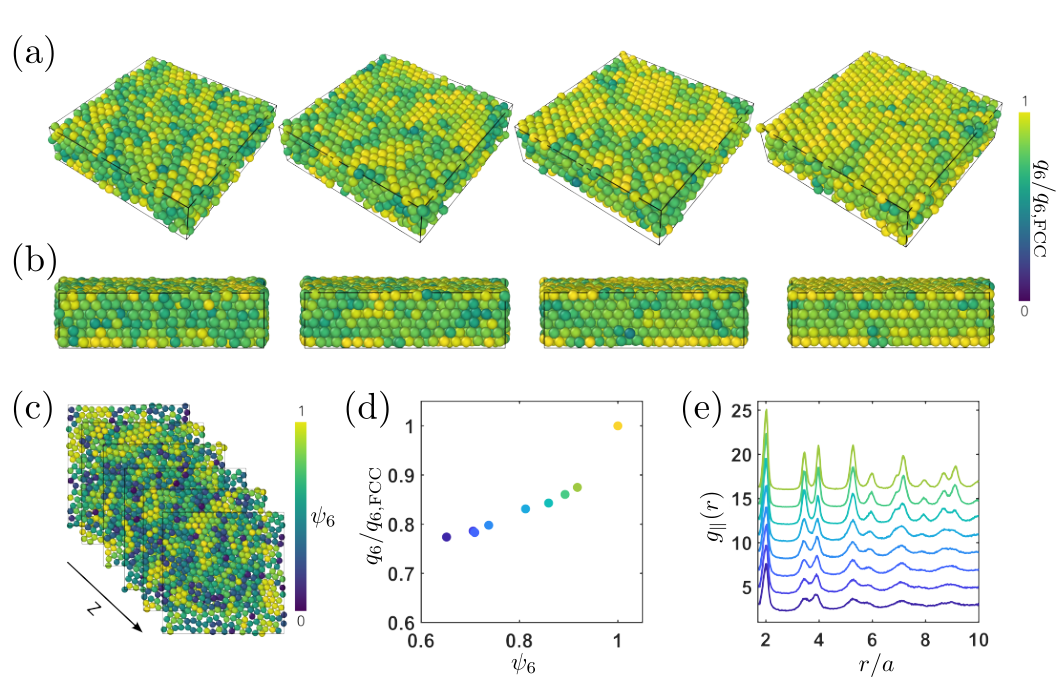}
    \caption{(a) Assembled SLs of six NP layers, formed by increasingly longer simulated compression times (left to right), exhibit enhanced structural order. Particles are colored by the value of their local 3D bond-orientational order parameter, $q_6$. 
    (b) Side views visualized using Ovito. \cite{ovito} 
    (c) Illustration of 2D layer binning to analyze intralayer structural order within a six-layer assembly, with particles colored by the value of their 2D hexatic bond-orientational order parameter, $\psi_6$. 
    (d) Normalized 3D bond-orientational order parameter ($q_6 / q_{6, {\rm FCC}} $) versus average 2D intralayer bond-orientational order parameter ($\psi_6$). 
    (e) Intralayer radial distribution function versus interparticle separation normalized by thermodynamic radius ($r/a$), as a function of increasing $\psi_6$ (bottom to top). Data arbitrarily offset for clarity.}
    \label{fig:Fig3}
\end{figure}

Assemblies formed from fast compression rates contain many NPs with $q_6$ values significantly lower than the average for a perfectly ordered six-layer FCC SL, $q_{6,{\rm FCC}}=0.581$ [Figures \ref{fig:Fig3}(a) and (b)]. Applying lower compression rates yields NP layers with increasingly larger grains of near-perfectly ordered NPs [see Table S1]. As expected, there is a strong correlation between the average $q_6$ of a six-layer NP assembly and the in-plane hexatic order $\psi_6$ present in its layers [Figures \ref{fig:Fig3}(c) and (d)], behavior observed in other equilibrium and nonequilibrium colloidal assemblies.\cite{truskett2000,torquato2000random,kansal2002diversity,atkinson2014existence} 
Consistent with the bond-orientational order parameter analysis, the in-plane radial distribution function, $g_{\parallel}(r)$, exhibits a characteristic split-second peak that becomes more pronounced in assemblies formed with slower compression rates [Figure \ref{fig:Fig3}(e)], a signature of enhanced translational order in colloidal hard-sphere monolayers.\cite{truskett_1998}  

The first step in understanding how disorder impacts polaritonic coupling in plasmonic NP SLs is to examine the properties of perfectly ordered assemblies. We begin our analysis with Au NP SLs, which have been extensively studied by theory, simulation, and experiments.\cite{Mueller_2018, Mueller_2020, mueller2021surface, Vieira_2019,Arul_2022,garcia_ultrastrong}.
We consider Au NPs with an optical diameter of $30$~nm and an overall area fraction of $\phi_0 \simeq 0.75$ in each layer, which is achieved for a gap of $3$~nm between nearest neighbors in a perfectly arranged structure. 
As discussed in Model and Methods, optical calculations are carried out using SMUTHI and system parameters are provided in Table \ref{table:system_parameters}.

There are two modes of interaction between metallic NP SLs and light that become hybridized to form plasmon-polariton states. The first is the collective transverse plasmon, which is the collective manifestation of the localized surface plasmon resonance of metallic NPs modified by near-field coupling due to the highly coordinated SL structure. The second are Fabry-P{\'e}rot cavity modes that are resonances that arise due to interference of light reflected from the two planar interfaces at the top and bottom of the SL. To isolate the cavity modes for study without the plasmon resonances, we investigate perfect Au NP SLs where the plasmonic cores are replaced by a medium with dielectric constant equal to Au's corresponding $\varepsilon_{\infty}$ value (Table~\ref{table:system_parameters}). 
In the absence of plasmonic cores, the films are non-absorbing, and the transmittance approaches $1$ while the reflectance approaches $0$ when a cavity mode forms. To determine how the number of NP layers affects the number of cavity modes within an NP film, we calculated reflectance ($R$) spectra while varying the number of layers, $N_{\rm layer}$. The number of cavity modes for a given $N_{\rm layer}$ can be estimated by counting dips in the corresponding $R$ spectra [Figure~S1(a)].
For 30 nm Au NPs, the cavity modes begin to appear for $N_{\rm layer} \geq 2$ within an excitation frequency range of $10^2$ cm\textsuperscript{-1} to $10^5$ cm\textsuperscript{-1}, i.e., a wavelength range of $100$ {\textmu}m to $100$~nm.  As expected, the number of cavity modes increases approximately linearly with increasing $N_{\rm layer}$ [Figure~S1(b)]. 

\begin{figure}[h]
    \centering
    \includegraphics[width=\textwidth]{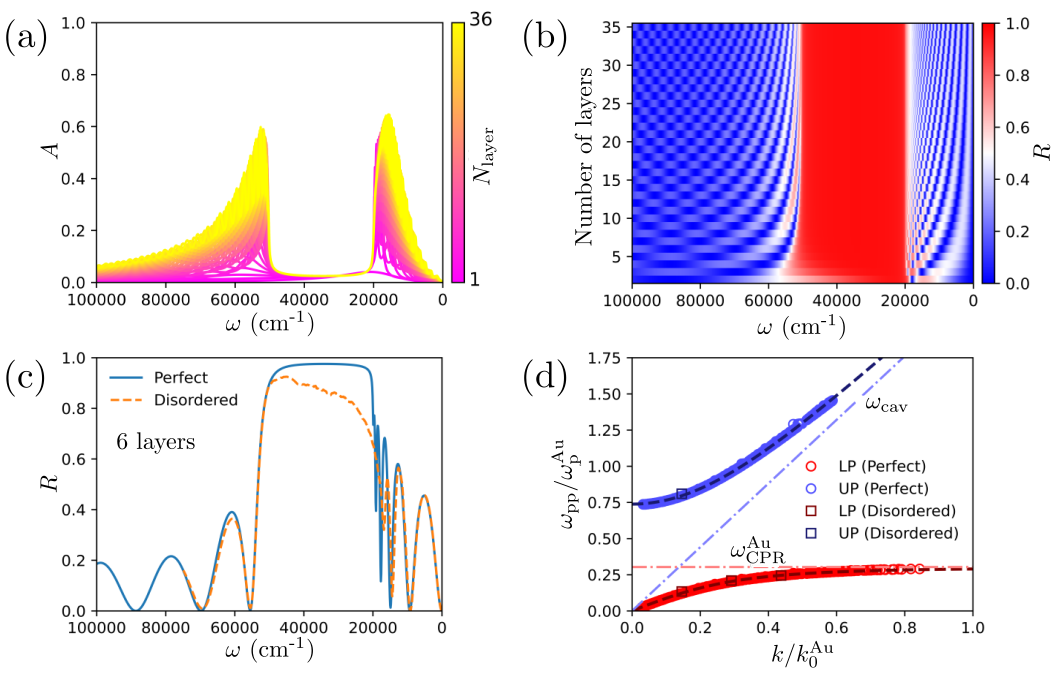}
    \caption{(a) Absorptance of perfectly ordered SLs consisting of $30$-nm diameter Au NPs with $3$-nm interparticle spacing from $1$ to $36$ layers (pink to yellow). 
    (b) Reflectance of the same system as a function of frequency and layer number.
    (c) Comparison between reflectance spectra of a perfect six-layer SL and a disordered six-layer structure, each consisting of 30-nm Au NPs. Due to the increasing computation time with increasing frequency, the latter is calculated up to the realization of two upper polariton states. 
    (d) Plasmon-polariton dispersion relation for the perfect SLs of varying layer numbers (circles) and six-layer disordered films (squares). Dashed lines are the Hopfield model fit. `LP' and `UP' stand for lower and upper polariton, respectively.}
    \label{fig:Fig4}
\end{figure}

When the plasmonic cores are included in the model for perfect Au NP SLs with $1$ to $36$ layers, the absorptance $A$ is enhanced and well-formed plasmon-polariton modes emerge with increasing $N_{\rm layer}$ and or film thickness [Figure \ref{fig:Fig4}(a)]. The plasmon-polariton modes appear as peaks in $A$, indicating strong absorption of photons at wavenumbers that satisfy the resonance condition given by eq~\ref{eq:pp_wl_resonance} [see schematic in Figure \ref{fig:Fig1}(a)].\cite{Vieira_2019, Mueller_2020, mueller2021surface}. 
High absorption correlates with reduced reflection, and polariton modes appear as dips in the reflectance ($R$) spectra [see Figure \ref{fig:Fig4}(b)]. 
Signatures of a polaritonic stop band in the frequency range $\approx 20000$ cm\textsuperscript{-1} $-$ $ 50000$ cm\textsuperscript{-1}, where photons are not absorbed but perfectly reflected, are evident in both the $A$ and $R$ spectra for $N_{\rm layer} \geq 4$. The stop band is incomplete (due to losses), but becomes sharper with increasing film thickness. 
Note that the Drude model---adopted here to provide a simplified description for the optical properties of metallic nanoparticles [eq~\ref{eq:Drude}]---allows us to model both the upper and lower polariton branches. In experiments on Au NP SLs, the upper polariton is obscured due to absorption by Au's interband transitions, which are not accounted for in the Drude model.\cite{Mueller_2020} 

To understand the effects of disorder on polaritonic coupling in NP multilayers, we compared $R$ computed for six-layered disordered configurations, as described in Model and Methods, to that for the perfect six-layer structure [Figure \ref{fig:Fig4}(c)]. To enable this comparison, we simulated disordered configurations with $361$ particles in each layer, totaling $2166$ particles in a six-layer structure, considering periodic images of the simulation box on the plane perpendicular to the direction of light propagation. The spectra were obtained by averaging over $20$ independently generated NP SL realizations with mean bond-orientational order parameters in a narrow window around $\psi_6=0.67$. 
These parameter choices were made to keep system sizes computationally feasible while also minimizing simulation artifacts due to lattice resonances, which can be excited in the implementation of SMUTHI with periodic boundaries even when the lateral dimension of the simulation box is several times the wavelength of interest \cite{theobald2021simulation, Theobald2022}. 
Interestingly, there were no noticeable differences in the polariton mode positions when comparing the perfect and disordered multilayers [Figure \ref{fig:Fig4}(c)]. The main distinctions were a lower reflectance magnitude for the disordered assembly, particularly in the stop band region, and the disappearance of a few of the higher-$j$ modes on the lower polariton branch near the stop band. 
Absorptance is higher for the disordered assembly than for the perfectly ordered one, particularly in the stop band, which is a signature of higher loss in the disordered system [see Figure S2(a)].

The polariton dispersion relations for the perfect and disordered multilayers can be extracted from the $R$ spectra. As described in Model and Methods, each mode number $j$ in the resonance condition given by eq~\ref{eq:pp_wl_resonance} corresponds to a polariton wavelength $\lambda_{\rm pp}$ and a pair of frequencies $\omega_{\rm pp}$, on the lower- and upper-polariton branches, respectively. The higher-$j$ modes on the lower-polariton branch and the lower-$j$ modes on the upper branch for a given layer can be challenging to identify and correctly index without high-frequency spectral resolution, as the differences between successive peaks and dips become increasingly small near the stop band [see Figure~S2(b)]. To help address this difficulty, the $j = 1$ mode position is first identified from the reflectance spectra for $N_{\rm layer} = 2$, where it is straightforward to locate, even though the stop band is not yet well-developed at this film thickness. 
As reported for the lower polariton branch in ref \onlinecite{Mueller_2020}, the mode position for a given $j$ shifts to lower frequencies as $N_{\rm layer}$ increases. This trend facilitates determining the mode number corresponding to the lowest clearly identifiable reflection dip for each $N_{\rm layer}$.
The upper and lower branches of the dispersion relation show a characteristic avoided crossing behavior [Figure \ref{fig:Fig4}(d)]; here, $\omega_{\rm pp}$ for Au NP multilayers is normalized by the Au plasma frequency, $\omega_{\rm p}^{\rm Au} = 68558$ cm\textsuperscript{-1}, and it is plotted versus polaritonic wavenumber $k=2\pi / \lambda_{\rm pp}$ normalized by $k_0$, the length of the first Brillouin zone in the $\Gamma L$ direction. 
For our Au SL, 
consisting of NPs of $d = 30$~nm diameter with $l_g = 3$~nm gaps between nearest neighbors, $k_0 = \sqrt{3}\pi/[\sqrt{2}(d+l_g)] \simeq 0.117$ nm\textsuperscript{-1}.
At large $k$, the upper polariton branch approaches the light line ($\omega_{\rm cav} = ck/n_{\rm eff}^{\rm Au}$) asymptotically. In contrast, the large $k$ asymptotic value of the lower polariton approaches the collective plasmon resonance $\omega_{\rm CPR}^{\rm Au}$.

The dispersion relation for the disordered six-layer structure shows excellent agreement with that of the perfectly ordered NP SL, leaving the upper and lower branches unaltered. 
We fit the Hopfield Hamiltonian dispersion relations given by eq~\ref{eq:HF_disp} to the dispersion data extracted for the perfect Au SL considering $n_{\rm eff}$, $\Omega_{\rm R}$, and $\omega_{\rm CPR}$ as free parameters. 
The best fit is obtained with $n_{\rm eff} = n_{\rm eff}^{\rm Au} = 1.23$, $\Omega_{\rm R} = \Omega_{\rm R}^{\rm Au} = 23.1 \times 10^3$ cm\textsuperscript{-1} ($\simeq 2.86$ eV), and $\omega_{\rm CPR} = \omega_{\rm CPR}^{\rm Au} = 20.8\times 10^3$ cm\textsuperscript{-1} ($\simeq 2.58$ eV). 
With these values, the normalized coupling strength, as defined in eq~\ref{eq:eta}, becomes $\eta=1.11$, which is in the deep strong coupling regime, following the nomenclature of ref \onlinecite{Mueller_2020}. This value signifies that disordered multilayers of metal NPs can exhibit the kind of extreme light-matter interactions established earlier for ordered plasmonic SLs.\cite{Mueller_2020}

\begin{figure}[hb!]
    \centering
    \includegraphics[width=0.5\textwidth]{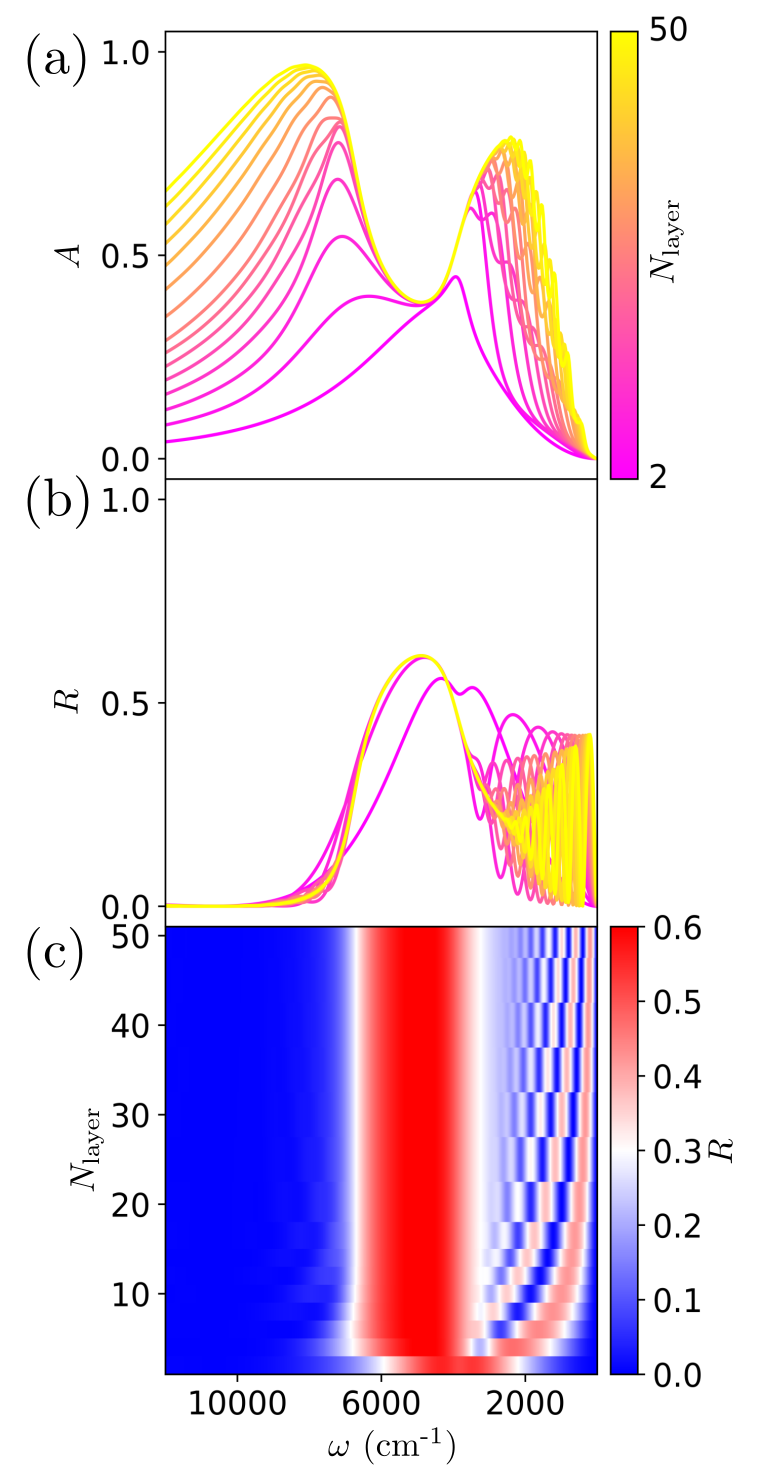}
    \caption{(a) Absorptance and (b) reflectance spectra of perfectly ordered SLs consisting of $90$-nm diameter ITO NPs with $9$-nm interparticle spacings from $2$ to $50$ layers (pink to yellow). 
    (c) Reflectance spectra of the same system as a function of frequency and layer number.}
    \label{fig:Fig5}
\end{figure}

Since the collective plasmon resonances of ITO NP assemblies are in the IR region, thicker films are required than for Au NP multilayers to generate hybridized plasmon-polariton states. One way to achieve thicker films is by increasing the number of layers compared to Au NP assemblies. However, this strategy requires large numbers of NPs, making systematic investigations by computer simulation infeasible. 
Instead, we consider larger, $d = 90$-nm diameter ITO particles. 
To keep the ratio $d/(d+l_g)$ and hence the area fraction $\phi_0$ in each layer the same as those in the case of Au NP films, we consider an interparticle gap length of $l_g = 9$~nm. 
Other system parameters are provided in Table \ref{table:system_parameters}. 
With these parameter choices, for ITO particles without plasmonic resonances, cavity modes would form for $N_{\rm layer} \geq 2$ within a frequency range of $\sim 100 - 25000$ cm\textsuperscript{-1} and their number increases approximately linearly with $N_{\rm layer}$ [Figure~S3].  
Similar to Au NP SLs, as the number of layers in a perfectly ordered ITO NP SL increases, $A$ increases as well as the number of polariton modes observed [Figures \ref{fig:Fig5}(a) and (b)]. 
The reflectance spectra, shown as a function of frequency and $N_{\rm layer}$ [Figure~\ref{fig:Fig5}(c)], indicate signatures of a polaritonic stop band from $\approx 4000 - 6000$ cm\textsuperscript{-1}, where the reflection is high and the absorptance is low. 

However, comparing the stop band signatures of ITO NP films to those of the Au NP assemblies, the former exhibits higher absorptance ($A$), lower reflectance ($R$), and less sharp lineshapes. These differences may be attributed in part to the higher intrinsic loss ($\gamma$) compared to the plasma frequency ($\omega_{\rm p}$) in the case of ITO NPs. 
To test this hypothesis, we simulated a model system consisting of NPs that have the same $\omega_{\rm p}$ as Au but $10$ times higher $\gamma$. 
The absorptance and reflectance of a perfectly ordered six-layered film of such NPs can be compared with that of the six-layered Au film (see Figure~S4). 
The sharpness of the stop band decreases with increasing $\gamma/\omega_{\rm p}$ ratio. In the reflectance spectra, the magnitude of the stop band diminishes, while it increases in the absorptance spectra, as expected, with increasing $\gamma$. 
Because of this, the upper polariton branches are formed in a higher, near-visible frequency range for ITO films and are not as prominent as the lower polariton modes [Figure \ref{fig:Fig5}]. 
Identical spectra over the complete simulation frequency range ($100 - 25000$ cm\textsuperscript{-1}) are shown in Figure~S5, where the faint ripple observed in both $A$ and $R$ traces the upper polariton branch, although its magnitude remains very low. 
Since the upper polariton branch appears in the near-visible range, we simulate the ITO films by considering contributions of both dipole and quadrupole NP moments. 
Figures~S6(a) and S6(b) feature a comparison of $A$ and $R$ obtained from dipole-only ($l=1$) and both dipole and quadrupole ($l=2$) calculations for $N_{\rm layer}=50$. 
The polariton frequencies from the $l=2$ simulations closely match those from the $l=1$ case, with only minor differences observed in the higher frequency region.

\begin{figure}[h]
    \centering
    \includegraphics[width=\textwidth]{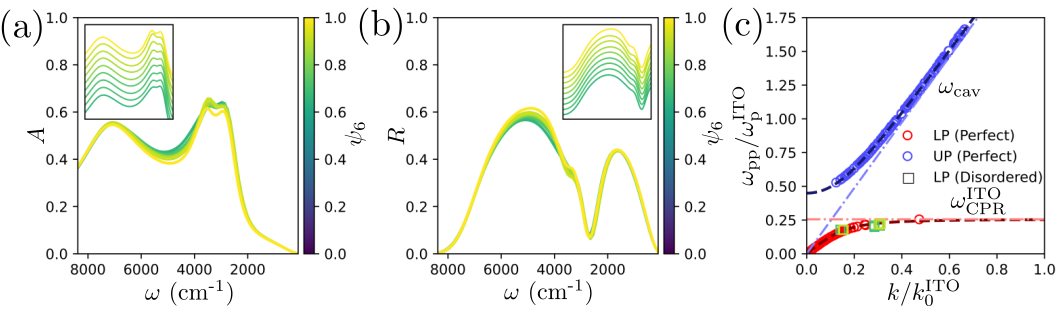}
    \caption{(a) Absorptance and (b) reflectance for six-layer films of $90$-nm diameter ITO NPs with varying average intralayer structural order, characterized by $\psi_6$. Lower polariton modes are labeled with the $j$ index of the standing wave. Insets are spectra offset for clarity. 
    (c) Plasmon-polariton dispersion for perfect ITO NP SLs of varying layer numbers (circles) and six-layer disordered films (squares). Dashed lines are Hopfield model fits. `LP' and `UP' stand for lower and upper polariton, respectively.}
    \label{fig:Fig6}
\end{figure}

To investigate the effect of structural heterogeneity on the plasmon polaritons of ITO NP multilayers,
we simulated six-layered disordered structures with a varying degree of disorder using SMUTHI.
We considered configurations, generated as described in Model and Methods, consisting of $d=90$-nm NPs, maintaining an area fraction of $\phi_0 \simeq 0.75$ in each layer.
Physical parameters, including the number of layers and particle diameter, were chosen to balance the computational cost in SMUTHI with the number of observed polariton modes. 
As for Au NP assemblies, we found that $\approx 300$ particles per layer were needed to describe the heterogeneous structures, so that lattice resonance artifacts imposed by the application of 2D periodic boundaries could be avoided and, thus, reliable trends could be produced. 
Each of our simulated six-layer disordered configurations contains a total of $2166$ NPs, with $361$ NP in each layer. 
For each value of $\psi_6$, we found that smooth and reliable spectra could be obtained from a single NP configuration, from which polariton positions could be accurately estimated. 
In Figures \ref{fig:Fig6}(a) and \ref{fig:Fig6}(b), we report $A$ and $R$ spectra of the disordered NP assemblies, respectively, as a function of frequency, restricting the range of interest within $\sim 100 - 8000$ cm\textsuperscript{-1}. 
Upon decreasing structural order ($\psi_6$), the stop-band reflectance decreases and the absorption increases. 
Hence, structural ordering contributes to effectively preventing light from entering the film; although none of the configurations acts as a total stop band, regardless of the microstructure. 
Upon closer inspection of the insets of Figures \ref{fig:Fig6}(a) and \ref{fig:Fig6}(b), it is apparent that structural heterogeneity also leads to broadened and less well-defined polariton modes, although the modal frequencies corresponding to $j=1$ do not vary significantly from those of the perfect SL [see Figure S7 (a)], whereas the estimated $j=2$ mode frequencies show stronger dependence on $\psi_6$ [see Figure S7(b)]. 
However, these modes are relatively broadened, making it challenging to estimate their exact positions and, consequently, to assess whether they follow a reliable trend. 
Importantly, even for the most disordered film ($\psi_6 = 0.65$), no modes vanish. 

This defect tolerance is further highlighted by the polariton dispersion relations computed from the reflectance spectra [see Figure S8]. 
The data for disordered configurations [Figure~\ref{fig:Fig6}(c)], irrespective of the degree of structural order, show good agreement with the data for the perfect SL, as in the case of Au NP films. 
For disordered ITO NP films, distinct reflection dips were identified for the lower-polariton branch only within the frequency range of interest. Furthermore, the difficulty in estimating the lower polariton mode frequencies corresponding to $j=2$ might have contributed to the minor deviations between their locations and the mode positions of the perfect ITO NP SLs. 

To estimate the strength of the plasmon-polariton coupling, we fit the Hopfield model to the dispersion data obtained for perfect ITO NP SLs treating $n_{\rm eff}$, $\Omega_{\rm R}$, and $\omega_{\rm CPR}$ as free parameters. 
We obtained the best fit with $n_{\rm eff} = n_{\rm eff}^{\rm ITO} = 1.70$, $\Omega_{\rm R} = \Omega_{\rm R}^{\rm ITO} = 2.77 \times 10^3$ cm\textsuperscript{-1} ($\simeq 0.34$ eV), and $\omega_{\rm CPR} = \omega_{\rm CPR}^{\rm ITO} = 3.83 \times 10^3$ cm\textsuperscript{-1} ($\simeq 0.47$ eV). 
The normalized coupling strength, in this case, was $\eta = \Omega_{\rm R}^{\rm ITO} / \omega_{\rm CPR}^{\rm ITO} \simeq 0.72$. 
The plasmon-polariton dispersion relations, extracted from the far-field spectra obtained with $l=2$ simulations, are shown in Figure~S6(c). 
The value of $\eta$ in this case is $0.73$, which is consistent with the value obtained for the $l=1$ simulations. 
Following the nomenclature of ref~\onlinecite{Mueller_2020}, ITO NP assemblies---whether disordered or perfectly ordered---fall into the category of ultrastrong coupling, whereas Au NP assemblies belong to the deep strong coupling regime, as discussed earlier. We note that recently reported ITO nanocrystal multilayer assemblies with disorder integrated into an open cavity structure also exhibited ultrastrong coupling.\cite{chang2025ultrastrong} 
See Table S2 for a list of the fitting parameters and direct comparisons with the corresponding values for the dipole-only model and the Au NP SLs.

We also estimate the normalized coupling strength, $\eta_{\rm Th}$, based on the theoretical framework for dipole plasmons presented in the Model and Methods section. 
The factor $f(k)$ is estimated for both Au SL and ITO SL using eq~\ref{eq:omega_CPR_micro}, where $\omega_{\rm CPR} = \omega_{\rm CPR}^{\rm Sim}$ is the extinction peak frequency of a monolayer (see Figure~S9) and $\omega_{\rm NP}$ is determined from eq~\ref{eq:omega_NP_Mie}. 
To estimate the dielectric constant $\varepsilon_{d}$, which quantifies the effect of the screening by the bound charges in a dense collection of NPs, we use eq~\ref{eq:neff}, with $n_{\rm eff}$ extracted by fitting the Hopfield model to the dispersion data. 
Assuming $f(k)$ remains approximately constant over the wavevector range of interest, we then calculate $\eta_{\rm Th}$ using eq~\ref{eq:eta_Th}. 
A comparison between $\eta_{\rm Th}$ and the fitted values of $\eta$, along with other related quantities, is presented in Table \ref{table:compare}. 
The calculated values of $\omega_{\rm NP}$ match the extinction peak position of a single NP for both Au and ITO (see Figure~S9). 
The values of $\omega_{\rm CPR}^{\rm Sim}$ for both Au SL and ITO SL agree with the values obtained from the dispersion relation fits. 
For Au SL, the value of $\eta$, obtained from the dispersion fit, is within $12\%$ of $\eta_{\rm Th}$, whereas this difference is only $4\%$ for ITO SL. 
The values of $\varepsilon_d$ are very close to $\varepsilon_{\infty}$ for both Au and ITO, where $\varepsilon_{\infty} = 1$ for Au and $\varepsilon_{\infty} = 4$ for ITO. 
We repeated the calculations with $\varepsilon_{\infty}$ in place of $\varepsilon_d$ in eqs~\ref{eq:g} and \ref{eq:eta_Th}, and the values of $\eta_{\rm Th}$ (Table S3) remained largely consistent with $\eta$ obtained from the dispersion fits. 

\begin{table}[h]
    \centering
    \begin{tabular}{|m{3.2cm} m{3cm} m{3cm}|}
    \hline
    Quantities                      & Au film   & ITO film  \\
    \hline
    $\epsilon_{d}$                  & 1.17         & 3.69         \\
    $\omega_{\rm NP}$ ($\times 10^3$ cm\textsuperscript{-1}) & $30.7$      & $5.30$       \\
    $\omega_{\rm CPR}^{\rm Sim}$ ($\times 10^3$ cm\textsuperscript{-1}) & $21.6$      & $4.01$       \\
    $f(k)$                          & -1.04         & -1.32         \\
    $\eta_{\rm Th}$                 & 0.99          & 0.75          \\
    $\eta$ (from fit)               & 1.11          & 0.72          \\
    $\tilde{\eta}$                  & 1.29          & 1.10          \\
    \hline
    \end{tabular}
    \caption{Quantities related to the theoretical estimation of the normalized coupling strength, $\eta_{\rm Th}$, for both Au SL and ITO SL.}
    \label{table:compare}
\end{table}

The oscillator strength $G$, as described in eq~\ref{eq:eps_eff_pp}, 
may be a useful parameter for comparing metamaterials because of its connection to the normalized coupling strength \cite{Mueller_2020}. 
Using the relationship $G=4 n_{\rm eff}^2 \eta^2$, we find $G \simeq 7.45$ for Au and $6.02$ for ITO NP films, respectively, where $n_{\rm eff}$ is obtained from the dispersion fit. 
One can alternatively use $\varepsilon_{\infty}$ instead of $\varepsilon_d$ in eq~\ref{eq:neff} to estimate $n_{\rm eff}$ and hence $G$ from $\eta$, which results in $G$ values of $6.90$ and $6.34$ for Au and ITO NP assemblies, respectively. 
By either approach, $G$ for ITO NP assemblies is predicted to be smaller than for Au NP films, but of a similar magnitude. For context, as noted in ref~\onlinecite{Mueller_2020}, these values of $G$ are three to five orders of magnitude higher than those typically observed for excitons in semiconductors.\cite{Yu1996,mihailovic2009optical}
They are, however, smaller than the maximum value ($G=16.6$) previously reported for Au NP SLs with higher layer area fractions, and hence stronger plasmonic near-field coupling, than those studied here.\cite{Mueller_2020} 
Additionally, we estimate $\tilde{\eta}$ using $\omega_{\rm CPR}^{\rm Sim}$ in eq \ref{eq:eta_tilde} for Au and ITO films studied here. 
The values are $1.29$ for Au and $1.10$ and ITO films, when $n_{\rm eff}$ obtained from the dispersion fit is used. 
Alternatively, using $\varepsilon_{\infty}$ instead of $\varepsilon_{d}$ in eq \ref{eq:neff} to estimate $n_{\rm eff}$ results in $\tilde{\eta}$ values of $1.34$ and $1.08$ for Au and ITO films, respectively. 
In both cases, $\tilde{\eta}$ is lower for ITO than the Au films, consistent with the trend observed for $\eta$, although $\tilde{\eta}$ overestimates the latter. 

To understand near-field coupling in perfect NP SL films when excited at a plasmon-polariton mode, we compute the local electric field intensities. 
Two orthogonal polarizations of the incident field (Figure~S10) result in distinct near-field intensity profiles in perfect structures.  
Within each layer ($xy$-plane) of a perfect SL, the NPs form a hexagonal arrangement, with nearest neighbors aligned along the $x$-axis. 
In the EY configuration [Figure~S10(a)], the electric field vector ($\vec{E}$) of the incident light oscillates along the $y$-axis, whereas, in the EX configuration [Figure~S10(b)], $\vec{E}$ oscillates along the $x$-axis, i.e., parallel to the chain of nearest-neighbor particles.
For a given mode number $j$, the local intensity is notably higher (`hot spots') in the EX configuration than in the EY configuration (Figure~S11). The $yz$ plane used here for visualization intersects the midpoint between two such particles, where the field concentration, or hot-spot intensity, is maximized. 
When a polariton mode on the upper polariton branch is excited, the NPs are more transparent and the field fully penetrates their volume, resulting in a diffuse near-field intensity distribution and reduced hot-spot intensities compared to the lower polariton branch (Figures~S12, S13, S14, and S15).  
The maximum near-field intensity in the upper branch is higher in the Au NP SL (Figures~S12 and S14) than in the ITO NP assembly (Figures~S13 and S15). Within the stop band, the light is predominantly reflected from the top layer, leaving the deeper regions of the NP film dark (Figures~S12, S13, S14, and S15).

\begin{figure}[h]
    \centering
    \includegraphics[width=\textwidth]{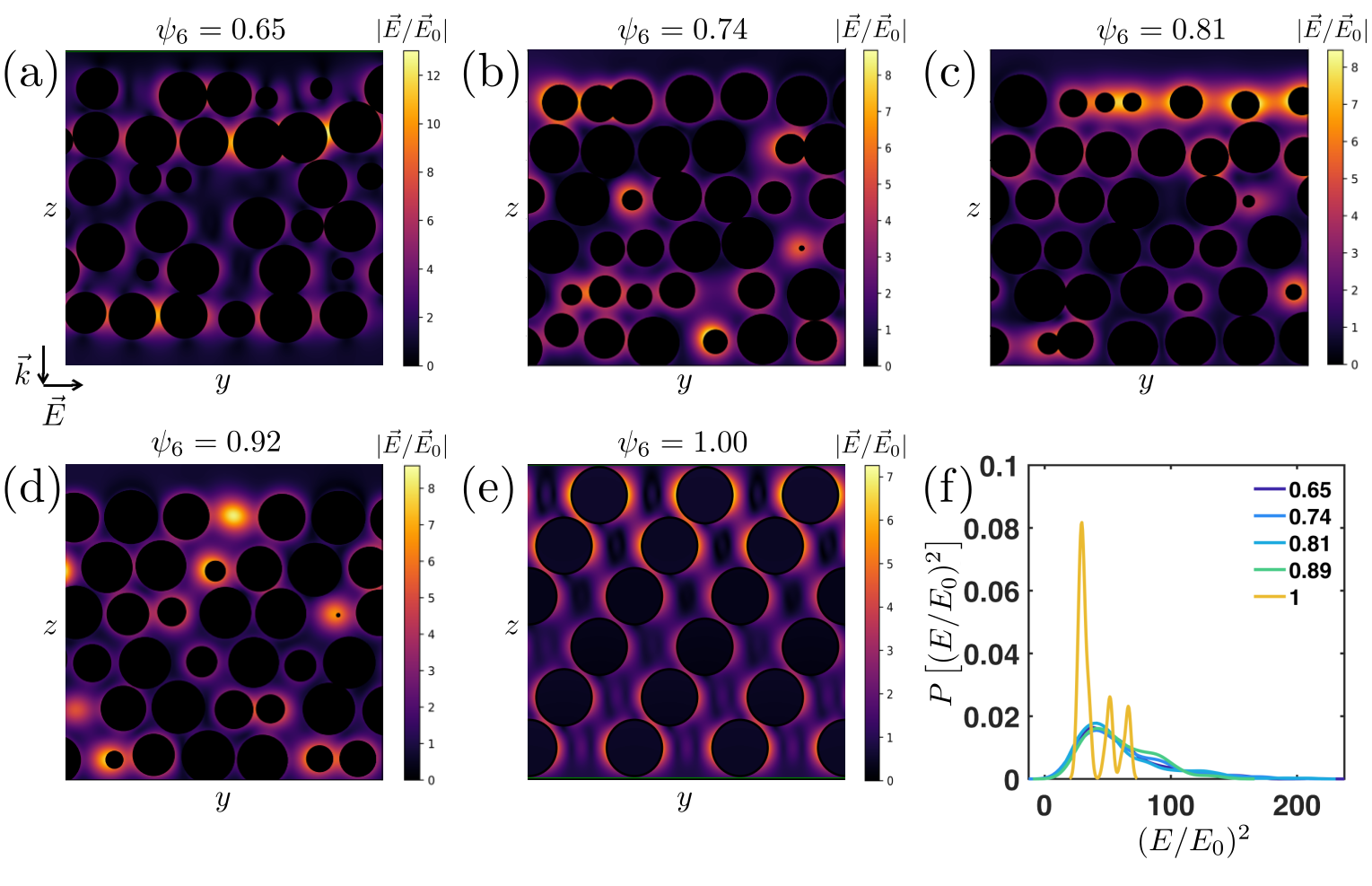}
    \caption{Near-field intensity maps in the $yz$ plane of a layer for six-layered films of $90$-nm ITO NPs with average $\psi_6$ of (a) 0.65, (b) 0.74, (c) 0.81, (d) 0.92, and (e) 1.00 (perfect SL). Particles located out of the image plane appear smaller due to projection effects. 
    (f) Probability distribution of hot spots in films of varying $\psi_6$. All field quantities are calculated at the $j = 1$ lower polariton frequency.
    }
    \label{fig:Fig7}
\end{figure}

Near-field intensity maps calculated at the $j = 1$ lower polariton frequency for ITO NP films with different $\psi_6$ values highlight the effects of structural disorder [Figures~\ref{fig:Fig7}(a)-(e)]. 
Here, $\vec{E}$ is polarized along the $y$-axis, corresponding to the EY configuration for the perfectly ordered structure ($\psi_6 = 1$). 
For clarity, each plot displays a section of the $yz$-plane that passes through the center of the simulation box. 
Despite the disorder, the field is predominantly concentrated in the interparticle gaps along chains of closely spaced NPs aligned with the incident field polarization direction, resulting in regions of intense hot spots (see also Figure~S11), as previously predicted by MPM simulations of disordered plasmonic NP monolayers\cite{green_2024}.  
More disorder leads to a broader distribution of hot spots, with intensities reaching up to $\sim 4$ times higher than in the perfect SL, as is evident from the probability distribution of the electric field intensity ($E^2$), normalized to that of the incident field ($E_0^2$), for configurations with varying $\psi_6$ values [Figure \ref{fig:Fig7}(f)]. The enhancement in the maximum local field intensity in disordered structures arises due to their significant fraction of particles that are closer than the interparticle spacing in the perfectly ordered structure of the same area fraction. Previous studies have established a strong correlation between near-field intensity distributions with a long, high-energy intensity tail and broadened far-field optical response in disordered plasmonic assemblies.\cite{green_2024,Sherman_PlasmonRuler_2023,sherman2024review} The far-field polaritonic mode broadening and near-field distributions reported here suggest that a similar mechanism underlies the optical responses of disordered plasmonic NP films. With both high near-field enhancements and a high density of hot spots present in a three-dimensional film, disordered plasmonic NP assemblies offer a promising alternative to perfect SLs for surface-enhanced Raman scattering and surface-enhanced infrared absorption \cite{Mueller_2021, Arul_2022, green_2024,chang2025ultrastrong}.

\begin{figure}[h]
    \centering
    \includegraphics[width=\textwidth]{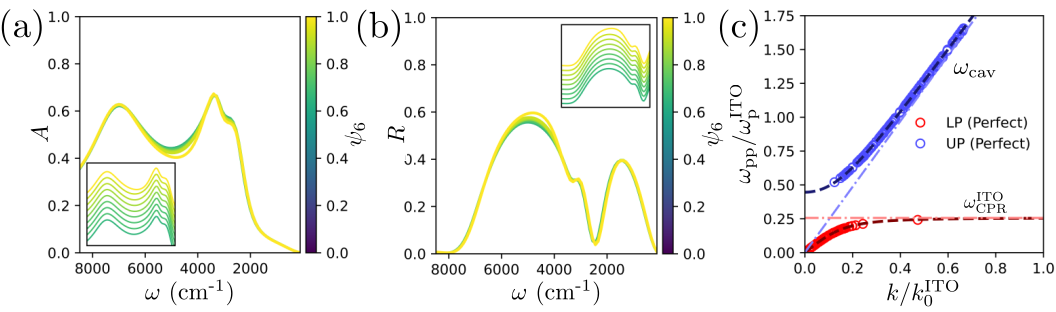}
    \caption{Far-field spectra of ITO NP films using MPM. (a) Absorptance and (b) reflectance calculated for six-layered films of varying $\psi_6$. 
    (c) Plasmon-polariton dispersion for perfect ITO SLs of varying $N_{\rm layer}$. Dashed lines are Hopfield model fits. `LP' and `UP' stand for lower and upper polariton, respectively.}
    \label{fig:Fig8}
\end{figure}

So far, we have presented optical properties predicted by T-matrix-based calculations performed using SMUTHI, where the NPs are modeled with dipolar plasmonic character. 
Here, we compare these results with those obtained from computationally efficient MPM calculations\cite{Sherman2023,Sherman_PlasmonRuler_2023,Kihoon_Meta_2023}, which account for mutual dipolar coupling within the quasistatic approximation. A detailed description of the implementation of MPM for layered materials can be found in ref~\onlinecite{Kihoon_Meta_2023}. First, MPM computations show good agreement with those carried out using SMUTHI for $A$ and $R$ spectra over a frequency range of $\sim 100-8000$ cm\textsuperscript{-1} for $90$-nm ITO NP multilayers exhibiting varying $\psi_6$ [Figures~\ref{fig:Fig8}(a) and (b)]. A more direct comparison was also carried out for absorptance and reflectance spectra of the perfectly ordered ITO film with $N_{\rm layer} = 50$ [Figures~S16(a) and (b)]. While the overall spectral features are qualitatively similar in this latter comparison, the MPM results show a more pronounced $j=2$ lower polariton mode and a slight redshift [Figures S16(c) and (d)] in both polariton branches relative to the computed branch locations obtained by SMUTHI. 
Though MPM predicts marginally stronger plasmonic coupling than obtained by the electrodynamic computations of SMUTHI, this difference is expected, given that MPM assumes every NP experiences the same electric field and thus does not accurately incorporate retardation effects.

We further extracted the dispersion relation from the MPM-predicted $R$ spectra of perfectly ordered structures with varying $N_{\rm layer}$ and fitted the Hopfield model to the predictions. 
Figure \ref{fig:Fig8}(c) demonstrates an excellent fit with parameters $n_{\rm eff} = 1.70$, $\Omega_{\rm R} = 2.74 \times 10^3$ cm\textsuperscript{-1} ($\simeq 0.34$ eV), and $\omega_{\rm CPR} = 3.84 \times 10^3$ cm\textsuperscript{-1} ($\simeq 0.48$ eV). 
The resulting normalized coupling strength is, in this case, $\eta=0.71$, which closely matches the value obtained from the SMUTHI-based analysis presented above. 
A comparison of the dispersion data obtained using both methods [Figure~S16(e)] reveals a close match, with both polariton branches appearing only slightly higher in energy in the case of SMUTHI. 

\section*{Conclusions}

Engineering light-matter interactions requires an understanding of the mechanisms that influence strong coupling. Assembling plasmonic NPs, with their tunable sizes and localized surface plasmon resonances, into SLs or multilayers with targeted interparticle spacings and number of layers, offers one promising strategy for designing optical response.\cite{Agrawal2018, Berry_Milliron_2024, Shevchenko2006,chang2023wavelength,chang2025ultrastrong}. 
Through electromagnetic simulation of the optical properties of 3D NP layered films with systematically varied structural disorder, we established that structural heterogeneity within the layers does not significantly modify the plasmon-polariton dispersion relation, which may help understand recently reported experimental observations for related NP assemblies.\cite{Arul_2022,chang2025ultrastrong} 
While forming low-defectivity assemblies is often considered an essential factor in design for optical properties, our calculations demonstrate that this in itself is not a stringent requirement for the emergence of plasmon-polariton modes or strong coupling.
However, we do find that structural heterogeneity generally leads to higher absorptance and lower reflectivity in the stop band compared to perfectly ordered structures. Structural disorder also leads to higher hot spot intensities, but broader near-field intensity distributions, than perfectly formed SLs. These considerations should be weighed against other factors when considering whether cheaper or simpler multilayer fabrication or assembly processes may be viable for achieving materials with the desired combination of properties.  
In this work, we did not consider effects from heterogeneities in particle size, shape, or dielectric properties, aspects that may also be interesting to incorporate into future studies. Understanding these and related aspects is part of an emerging focus in the field on the role that structural disorder can play in limiting or enhancing the optical response of materials.\cite{vynck2023light,sherman2024review,lalanne2025disordered,forster2010biomimetic,froufe2016role,froufe2017band,herkert2023influence}

This study also represents one of the first computational design investigations of ITO NP films, which are promising candidates for IR-active metamaterials.\cite{chang2024plasmonic} 
Doped metal oxides, such as ITO, have higher electron damping and lower carrier concentrations than noble metals, which limit near-field enhancement and increase spectral broadening.\cite{Agrawal2018} As shown here, structural disorder may further exacerbate losses in these assemblies, limiting their viability for some applications. Although the polaritonic coupling in ITO NP films is weaker than that observed in Au NP SLs, it still lies in the ultrastrong coupling regime. Moreover, ITO NP films benefit from the intrinsic tunability of their optical properties, which can be modified independently of their diameter through doping. This tunability expands the parameter space for enhancing coupling strength and stop-band characteristics, improving the overall polaritonic behavior. 
Comprehensive studies that systematically vary these parameters are essential for understanding the potential of ITO and other doped metal oxide NPs as viable building blocks for tunable light–matter interactions. 

Further investigations into plasmon-polaritons in NP assemblies are necessary to develop a more comprehensive design framework for achieving a desired optical response. One aspect that deserves more focus is the development of techniques for quantifying\cite{maher2024local} and controlling\cite{roach2022controlling,demirors2024tuning} structural disorder in nanoparticle assemblies.  
The selection of the electromagnetic simulation tool also plays a vital role in the accessible investigation areas for future studies. Here, we employed a dipole-only implementation of the T-matrix method, which enabled the analysis of six-layer disordered films at a moderate computational cost. While the T-matrix approach permits the selection of higher-order multipoles, detailed modeling with $\mathcal{O}(10^3)$ particles or more becomes prohibitively expensive. 
Simplifying assumptions should allow investigations of a more expansive parameter space. 
Based on its successful reproduction of polariton dispersion relations from SMUTHI, MPM could enable modeling of larger-scale assemblies at low computational cost in the quasistatic limit. 
MPM could be used to study NP films with increased layer numbers, as the quasistatic approximation does not appear to affect the resulting optical response significantly. However, a more systematic investigation of this latter point is warranted. The increased computational efficiency should also permit the efficient integration of optical property computations with inverse methods for materials design.\cite{kadulkar2022machine,liu2021tackling}

\begin{acknowledgement}

 This research was sponsored by the Army Research Office and was accomplished under Grant Number W911NF-23-1-0387. D.J.M. and T.M.T. acknowledge support from the Welch Foundation (F-1848 and F-1696).  The authors acknowledge the National Science Foundation through the Center for Dynamics and Control of Materials: an NSF Materials Research Science and Engineering Center (NSF MRSEC) under Cooperative Agreement DMR-2308817 and the Texas Advanced Computing Center (TACC) for providing computing resources. We thank Amos Egel, Dominik Theobald, and Zachary Sherman for helpful discussions.

\end{acknowledgement}

\begin{suppinfo}

Supporting Information is attached after References.
Additional plots supporting analysis details and tables containing information regarding Brownian dynamics simulation details and optical simulation results.

\end{suppinfo}


\providecommand{\latin}[1]{#1}
\makeatletter
\providecommand{\doi}
  {\begingroup\let\do\@makeother\dospecials
  \catcode`\{=1 \catcode`\}=2 \doi@aux}
\providecommand{\doi@aux}[1]{\endgroup\texttt{#1}}
\makeatother
\providecommand*\mcitethebibliography{\thebibliography}
\csname @ifundefined\endcsname{endmcitethebibliography}  {\let\endmcitethebibliography\endthebibliography}{}

\newpage

\begin{center}
    \textbf{\sffamily \LARGE Supporting Information: Plasmon Polaritons in Disordered NP Assemblies} \\
    {\sffamily \large Tanay Paul,\textsuperscript{\dag,\S} Allison M. Green,\textsuperscript{\dag,\S} Delia J. Milliron,\textsuperscript{*,\dag,\ddag} and Thomas~M.~Truskett\textsuperscript{*,\dag,\P}} \\
    \dag \textit{McKetta Department of Chemical Engineering, University of Texas at Austin, Austin, Texas 78712, United States} \\
    \ddag \textit{Department of Chemistry, University of Texas at Austin, Austin, Texas 78712, United States} \\
    \P \textit{Department of Physics, University of Texas at Austin, Austin, Texas 78712, United States} \\
    \S \textit{These authors contributed equally to this work.} \\
    {\sffamily E-mail: milliron@che.utexas.edu; truskett@che.utexas.edu}
\end{center}

\vspace{20mm} 
\textbf{Contents: 12 pages, 16 figures, 3 tables}

\newpage

\setcounter{figure}{0}
\renewcommand{\figurename}{Figure}
\renewcommand{\thefigure}{S\arabic{figure}}

\section{Supporting Figures}

\begin{figure}[htbp!]
     \centering
     \includegraphics[width=1.0\textwidth]{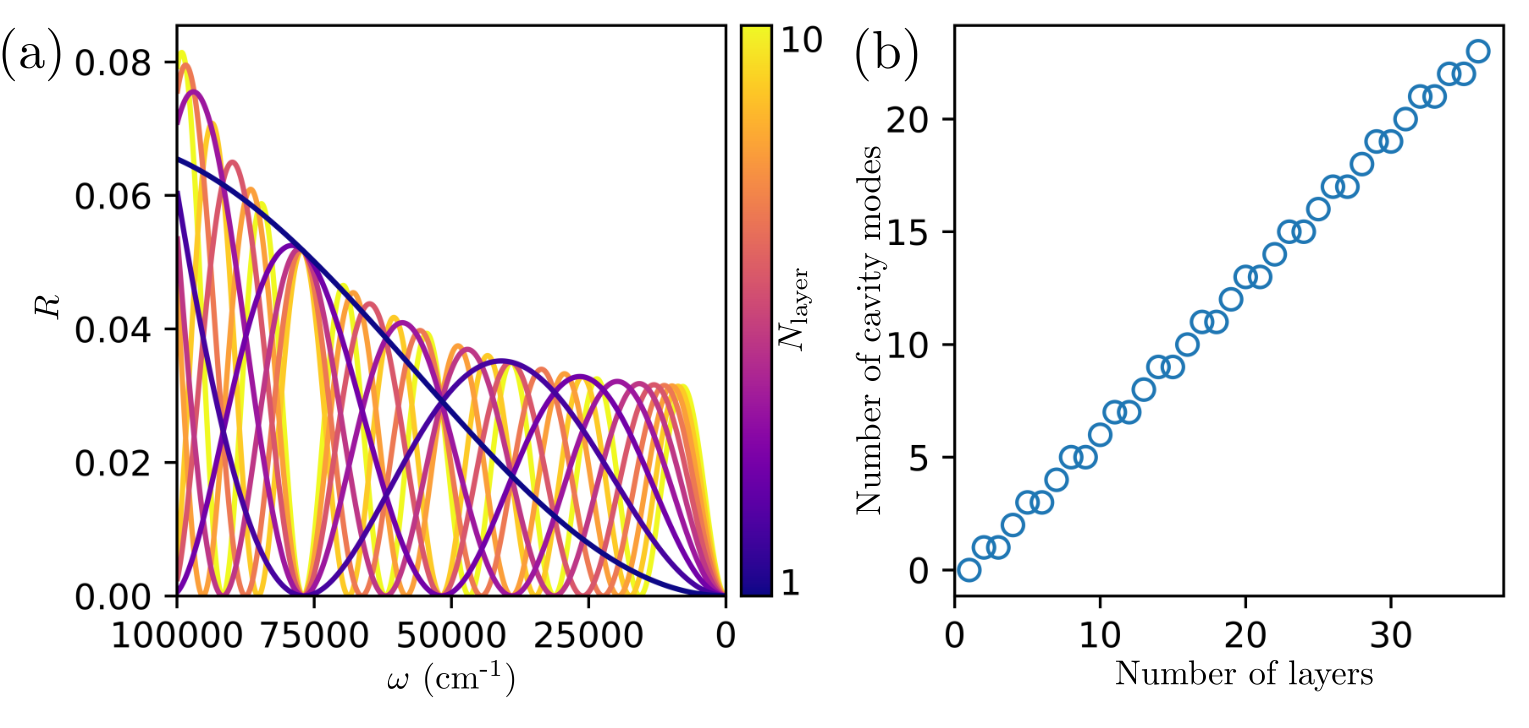}
     \caption{(a) Reflectance spectra for SLs of varying number of layers, $N_{\rm layer}$, where Au NPs are assigned dielectric constant $\varepsilon_{\infty}=1$, removing their plasmonic resonance. Reflectance dips represent Fabry-P{\'e}rot cavity mode frequencies. (b) Number of cavity modes as a function of $N_{\rm layer}$. }
     \label{SIfig:cavitymodes_nL_forAu}
\end{figure}

\begin{figure}[htbp!]
    \centering
    \includegraphics[width=1.0\textwidth]{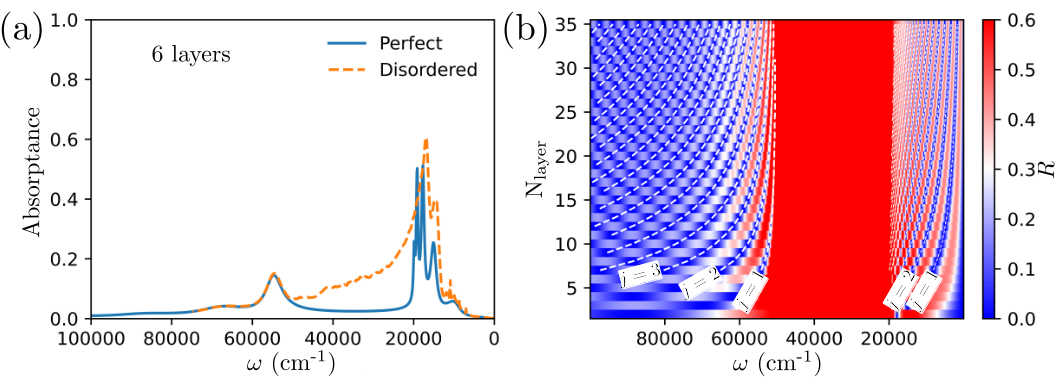}
    \caption{(a) Comparison between absorptance spectra of a perfect six-layer SL and a disordered six-layer Au NP film. (b) Reflectance of perfectly ordered Au NP SLs as a function of $N_{\rm layer}$ and wavenumber, $\omega$, with the constant-$j$ modes marked by dashed lines for both lower and upper polariton branches. The first few modes are labeled.}
    \label{SIfig:perfect_vs_disorder_AuSL}
\end{figure}

\begin{figure}[htbp!]
     \centering
     \includegraphics[width=1.0\textwidth]{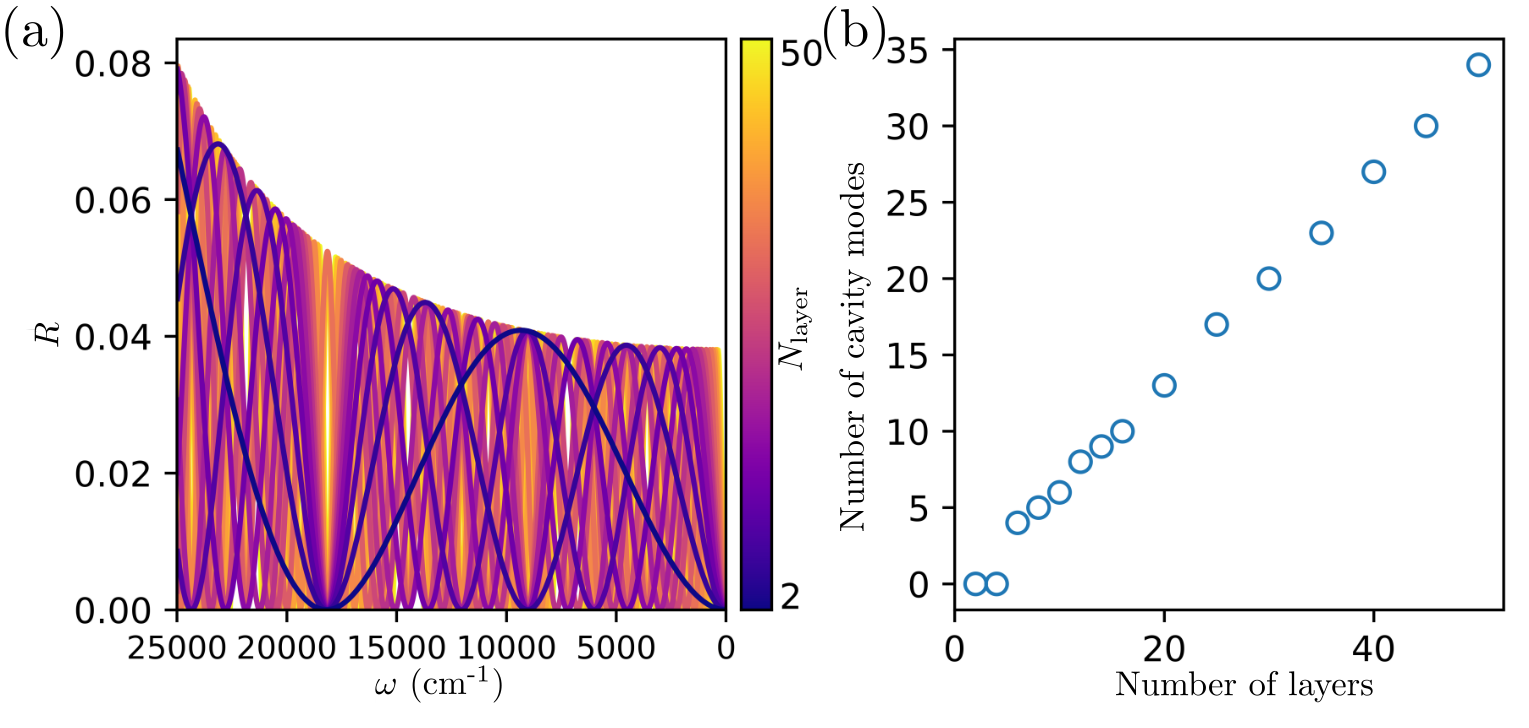}
     \caption{(a) Reflectance spectra for SLs of varying $N_{\rm layer}$, when ITO NPs are assigned a dielectric constant $\varepsilon_{\infty}=4$ to remove their plasmonic resonance. Reflectance dips represent the Fabry-P{\'e}rot cavity mode frequencies. (b) Number of cavity modes as a function of $N_{\rm layer}$. }
     \label{SIfig:cavitymodes_nL_forITO}
\end{figure}

\begin{figure}[htbp!]
     \centering
     \includegraphics[width=1.0\textwidth]{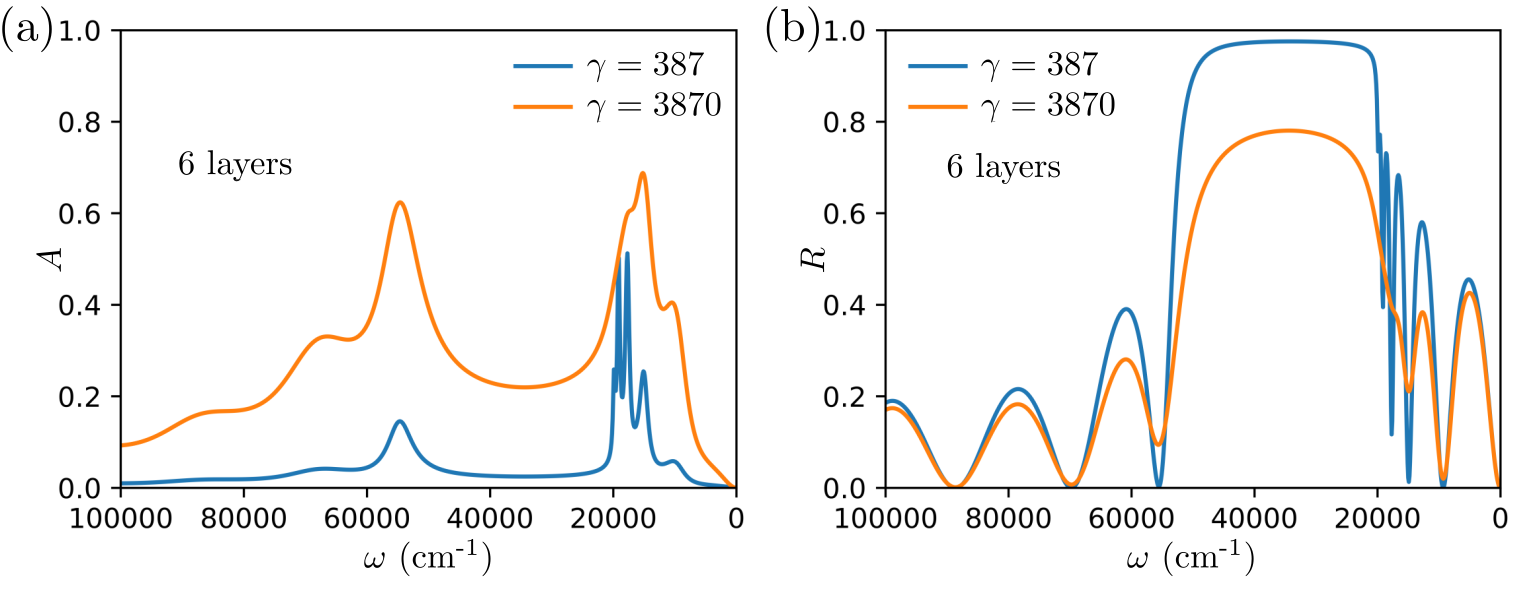}
     \caption{Comparison of (a) absorptance and (b) reflectance between two perfectly ordered six-layered films, one consisting of model Au NP and the other consisting of particles having the same plasma frequency ($\omega_{\rm p}$), but $10$ times higher damping, $\gamma$.}
     \label{SIfig:farfield_compare_gamma_AuSL}
\end{figure}

\begin{figure}[htbp!]
     \centering
     \includegraphics[width=1.0\textwidth]{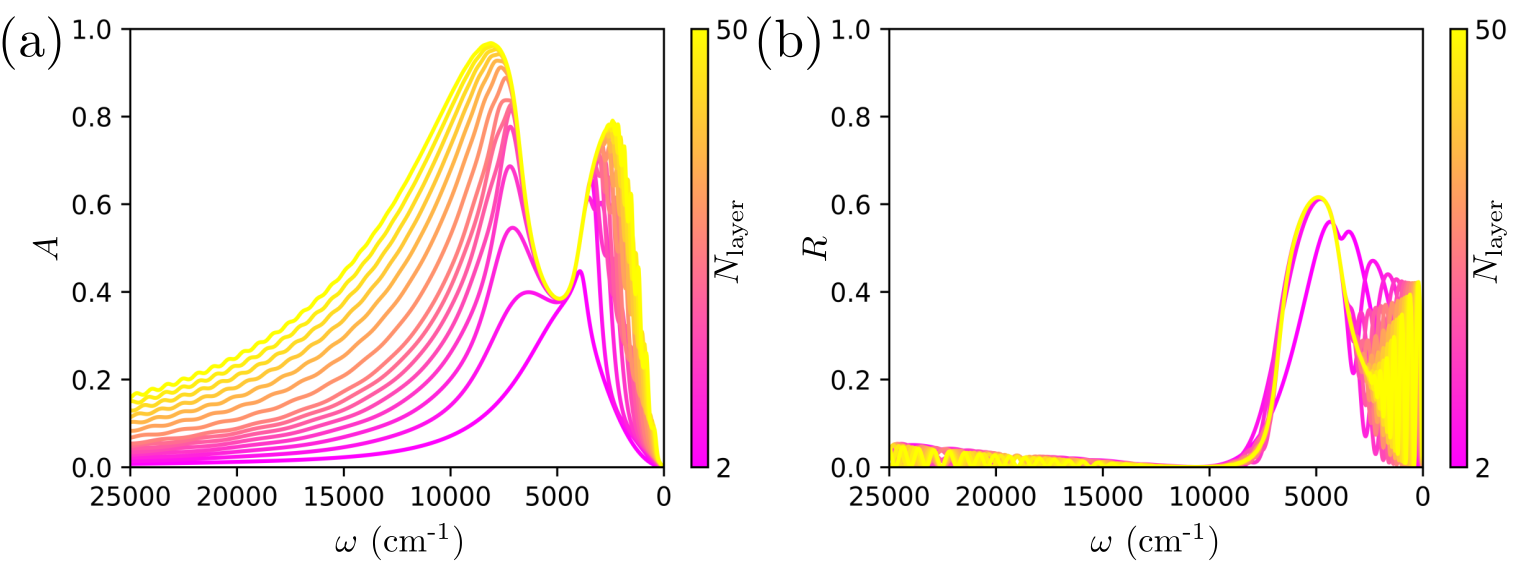}
     \caption{(a) Absorptance and (b) reflectance spectra for the full wavenumber range of interest that contains both lower and upper polariton branches for perfectly ordered ITO films.}
     \label{SIfig:far-field_ITOfilms}
\end{figure}

\begin{figure}[htbp!]
     \centering
     \includegraphics[width=1.0\textwidth]{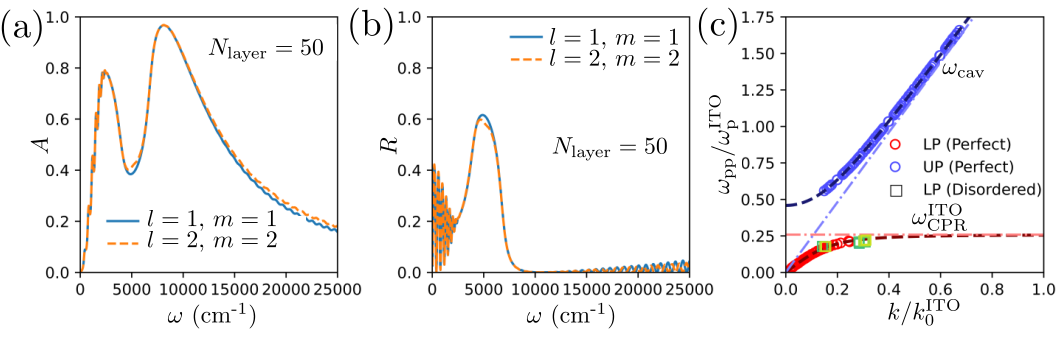}
     \caption{Comparisons of (a) absorptance and (b) reflectance calculated with a choice of $l=1, m=1$ and $l=2, m=2$ for ITO NP films. The first choice considers dipole plasmons only, whereas the second one considers quadrupolar in addition to dipolar contributions. (c) Dispersion relations for the $l=2, m=2$ case. The dashed lines are Hopfield model fits with parameters given in Table \ref{table:compare_fit_params}. The dispersion data for the six-layered disordered systems with varying degrees of disorder ($\psi_6$) are shown by squares. `LP' and `UP' stand for lower and upper polariton, respectively.}
     \label{SIfig:Smuthi_lcomp}
\end{figure}

\begin{figure}[htbp!]
     \centering
     \includegraphics[width=1.0\textwidth]{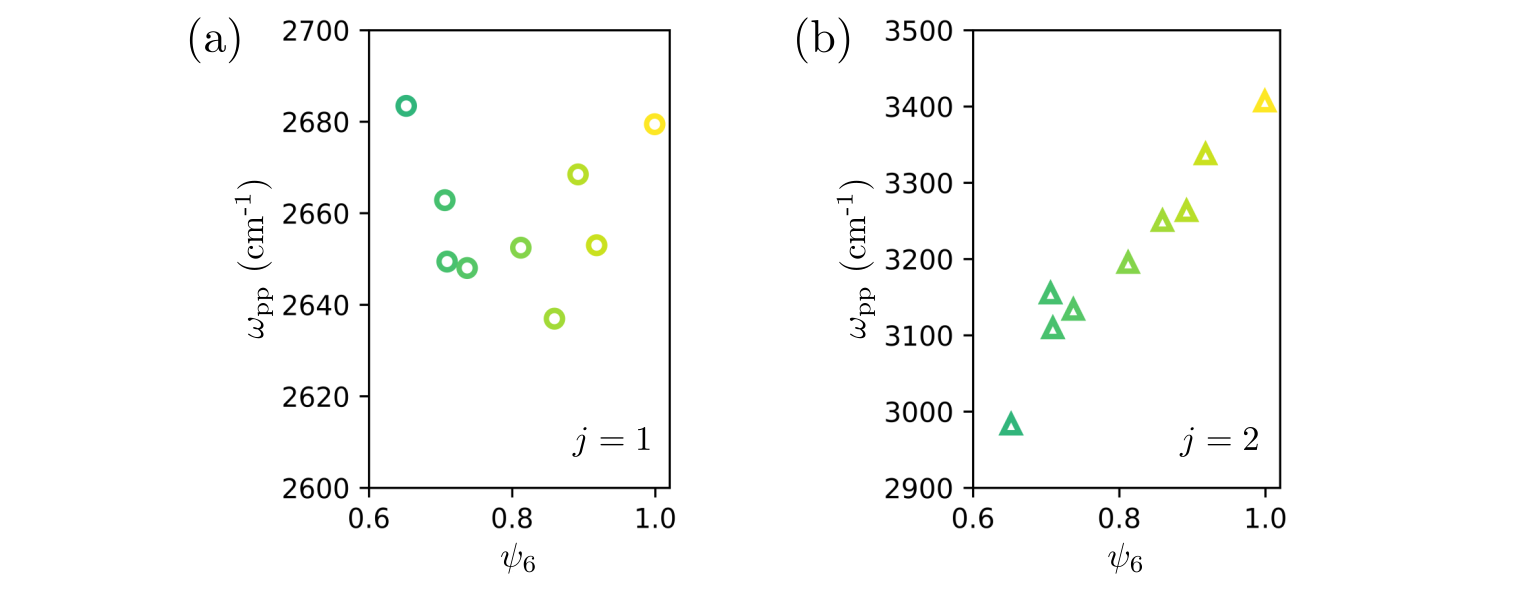}
     \caption{Estimated polariton peak reflectance frequencies for (a) $j=1$ and (b) $j=2$ for six-layer disordered ITO films with varying average intralayer structural order, characterized by $\psi_6$. The colors of the points vary from green to yellow as $\psi_6$ increases (see the colorbars in Figure 6).}
     \label{SIfig:polariton_modes_ITO_SL}
\end{figure}

\begin{figure}[htbp!]
    \centering
    \includegraphics[width=1.0\textwidth]{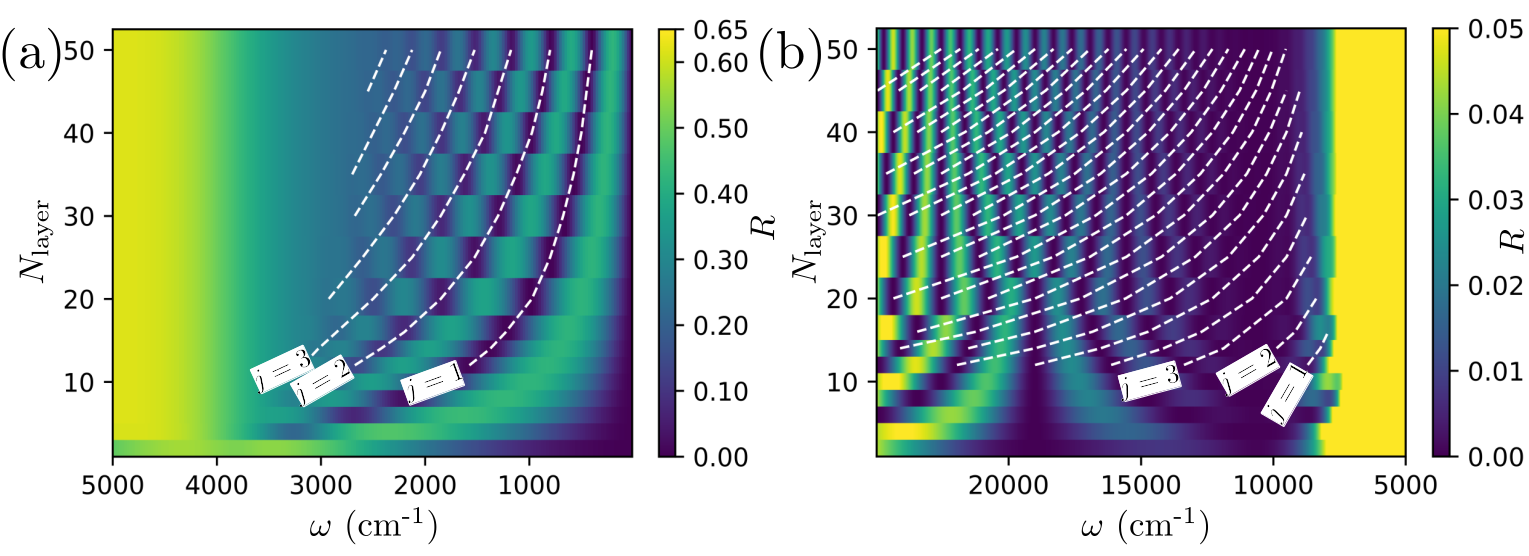}
    \caption{Reflectance of perfectly ordered ITO NP SLs as a function of $N_{\rm layer}$ and $\omega$ with the constant-$j$ modes marked by dashed lines for both lower and upper polariton branches. (a) Contains the $\omega$ range of the lower polariton branch. (b) Shows the upper polariton branch only for a clear visualization of the modes. The first few modes are labelled. The dashed lines are included to clarify the mode positions, which are challenging to resolve in a heat map due to the discrete nature of $N_{\rm layer}$.}
    \label{SIfig:R_vs_freq_vs_nL_j_marked_ISL}
\end{figure}

\begin{figure}[htbp!]
     \centering
     \includegraphics[width=1.0\textwidth]{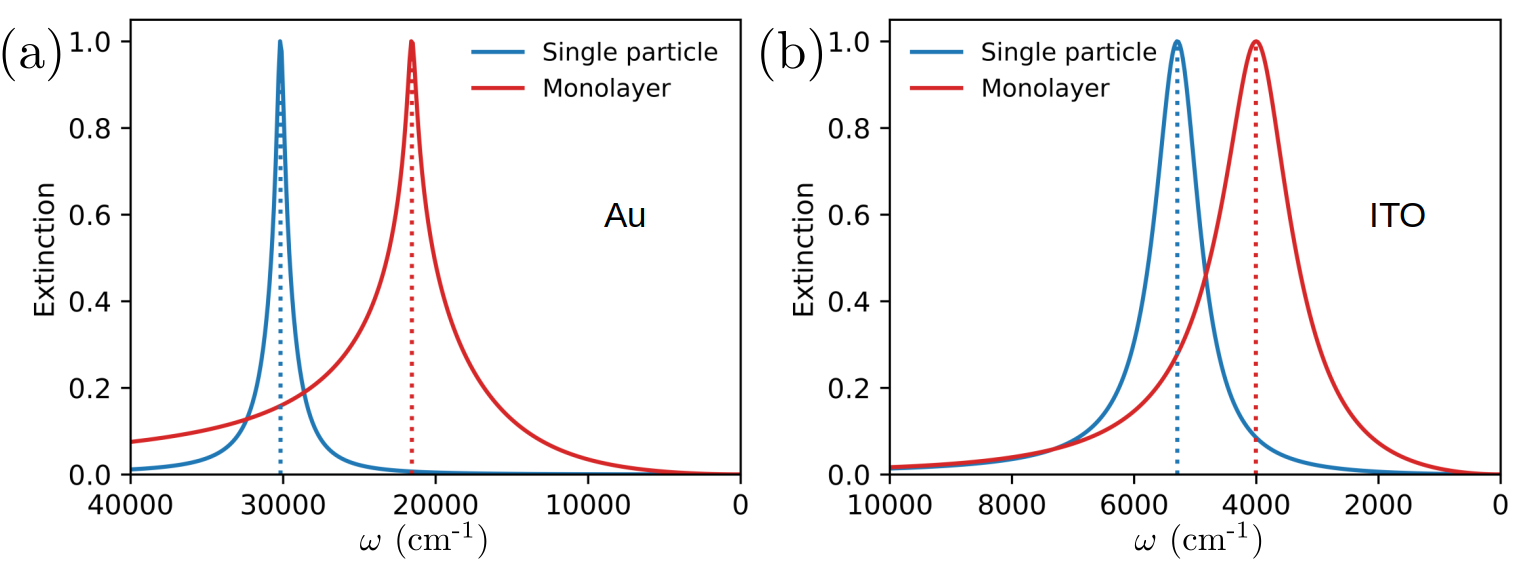}
     \caption{Plots of extinction, calculated as $-\log_{10}(T)$, where $T$ is the transmittance, and then normalized to $1.0$, for a single-particle and the perfectly ordered monolayer configurations consisting of (a) Au and (b) ITO NPs. The peak positions, marked by dotted lines, are obtained by fitting the data points near the peak with Gaussian functions. The single-particle extinction peak frequency is $\omega_{\rm NP}^{\rm Sim} = 30172.3$ cm\textsuperscript{-1} for Au and $5294.81$ cm\textsuperscript{-1} for ITO. These values match the values obtained from Eq. (19) and as reported in Table 2. The monolayer extinction peak frequency is described as the collective plasmon resonance (CPR) frequency, and the values are $\omega_{\rm CPR}^{\rm Sim} = 21561.3$ cm\textsuperscript{-1} for Au and $4009.03$ cm\textsuperscript{-1} (see Table 2).}
     \label{SIfig:single_particle_monolayer_extinctions}
\end{figure}

\begin{figure}[htbp!]
    \centering
    \includegraphics[width=1.0\textwidth]{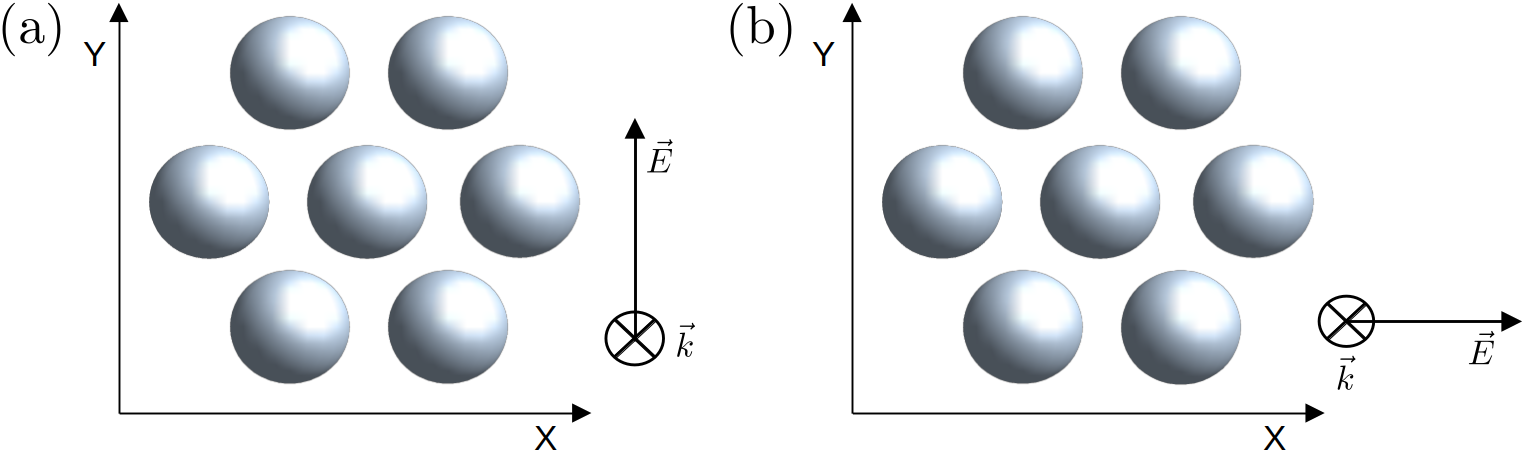}
    \caption{Two different configurations that are important for the near-field study in perfectly ordered structures. (a) The EY configuration: The electric field, $\vec{E}$, is perpendicular to the direction connecting nearest neighbour particles (the $y$ axis). (b) The EX configuration: $\vec{E}$ is parallel to the direction connecting the nearest neighbour particles (the $x$ axis). The wavevector, $\vec{k}$, is perpendicular to the $xy$ plane. }
    \label{SIfig:EY_EX_representations}
\end{figure}

\begin{figure}[htbp!]
    \centering
    \includegraphics[width=1.0\textwidth]{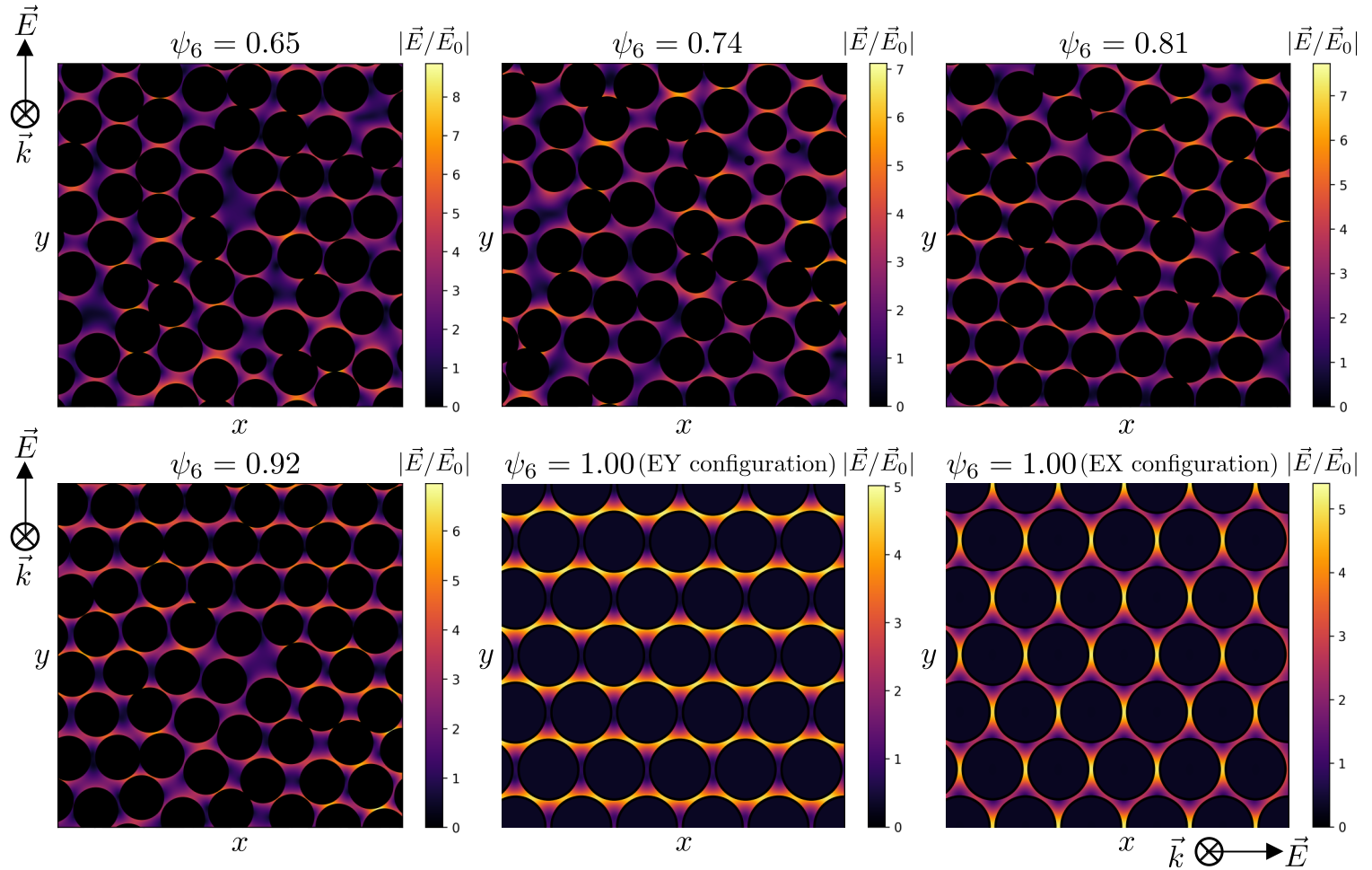}
    \caption{Near-field maps for ITO NP films of varying degrees of disorder computed at $j=1$ lower polariton mode frequency on one layer plane (the $xy$ plane), which is perpendicular to $\vec{k}$. Fields are calculated at the plane passing through the mean position of all the particles in the third layer of a six-layered film along the direction of $\vec{k}$. The $\psi_6$ values indicate the degree of disorder in the corresponding configuration, with $\psi_6 = 1.00$ denoting the perfectly ordered structure; please see Methods. Near-field maps for both EY and EX configurations are provided for the perfectly ordered structure. A section of the whole layer plane is shown for clarity in each case.}
    \label{SIfig:NF_xy_ISL}
\end{figure}

\begin{figure}[htbp!]
    \centering
    \includegraphics[width=1.0\textwidth]{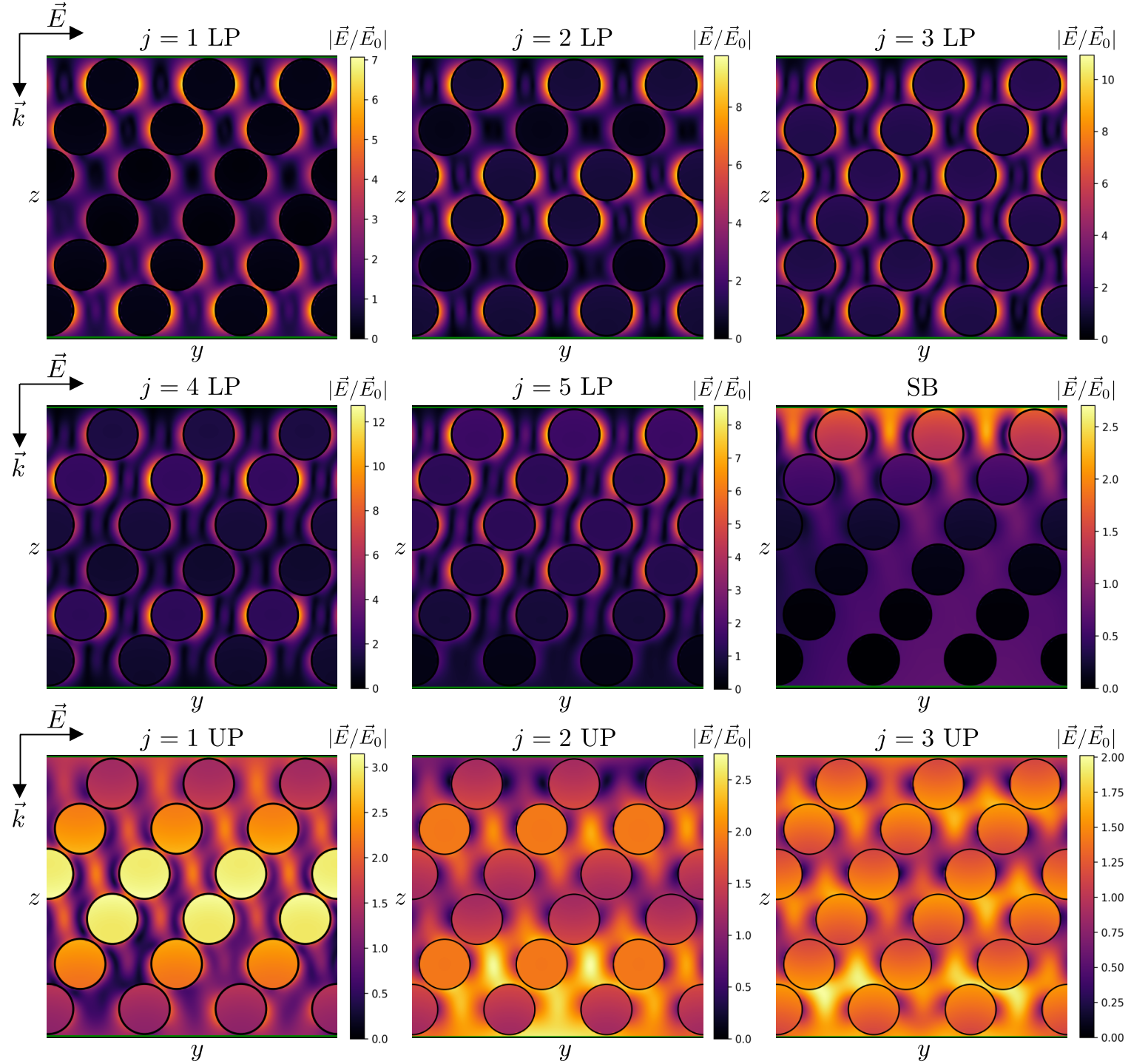}
    \caption{Near-field maps for perfectly ordered six-layered Au NP films at all observed $j$ modes corresponding to both lower polariton (LP) and upper polariton (UP) branches and at one frequency inside the stop band (SB). The presented field maps are computed on a plane that contains both $\vec{k}$ and $\vec{E}$, the $yz$ plane. The direction of the $\vec{E}$ corresponds to the EY configuration.}
    \label{SIfig:NF_TE_xz_ASL}
\end{figure}

\begin{figure}[htbp!]
    \centering
    \includegraphics[width=1.0\textwidth]{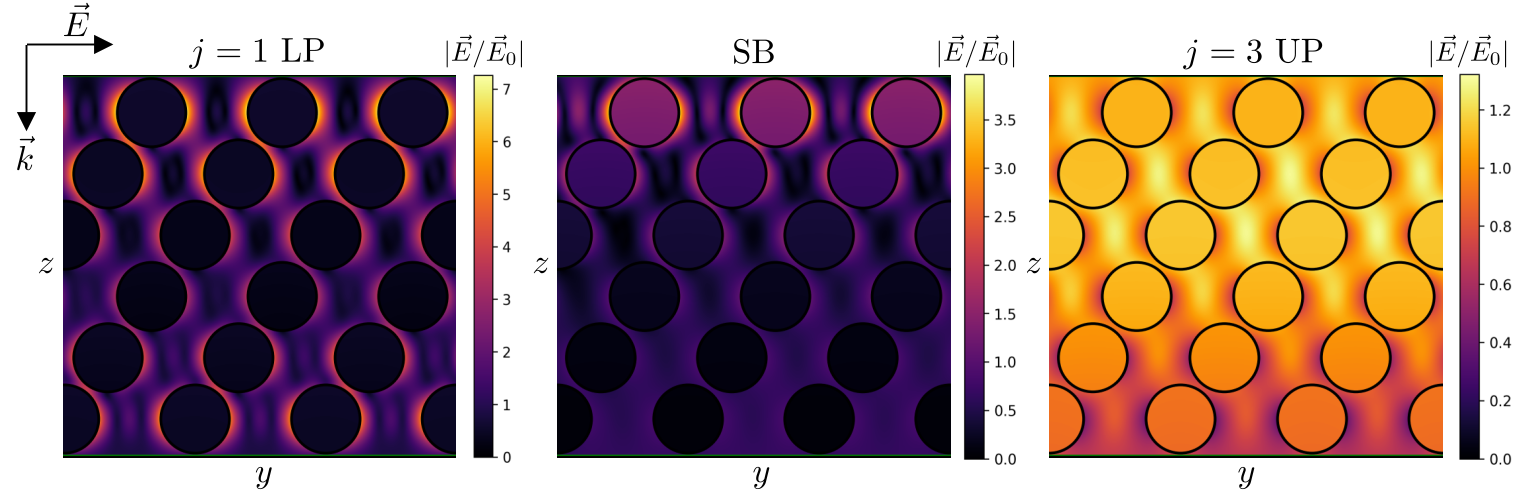}
    \caption{Near-field maps for perfectly ordered six-layered ITO films at all observed $j$ modes corresponding to both lower polariton (LP) and upper polariton (UP) branches and at one frequency inside the stop band (SB). The presented field maps are computed on a plane that contains both $\vec{k}$ and $\vec{E}$, the $yz$ plane. The direction of the $\vec{E}$ corresponds to the EY configuration.}
    \label{SIfig:NF_TE_xz_ISL}
\end{figure}

\begin{figure}[htbp!]
    \centering
    \includegraphics[width=1.0\textwidth]{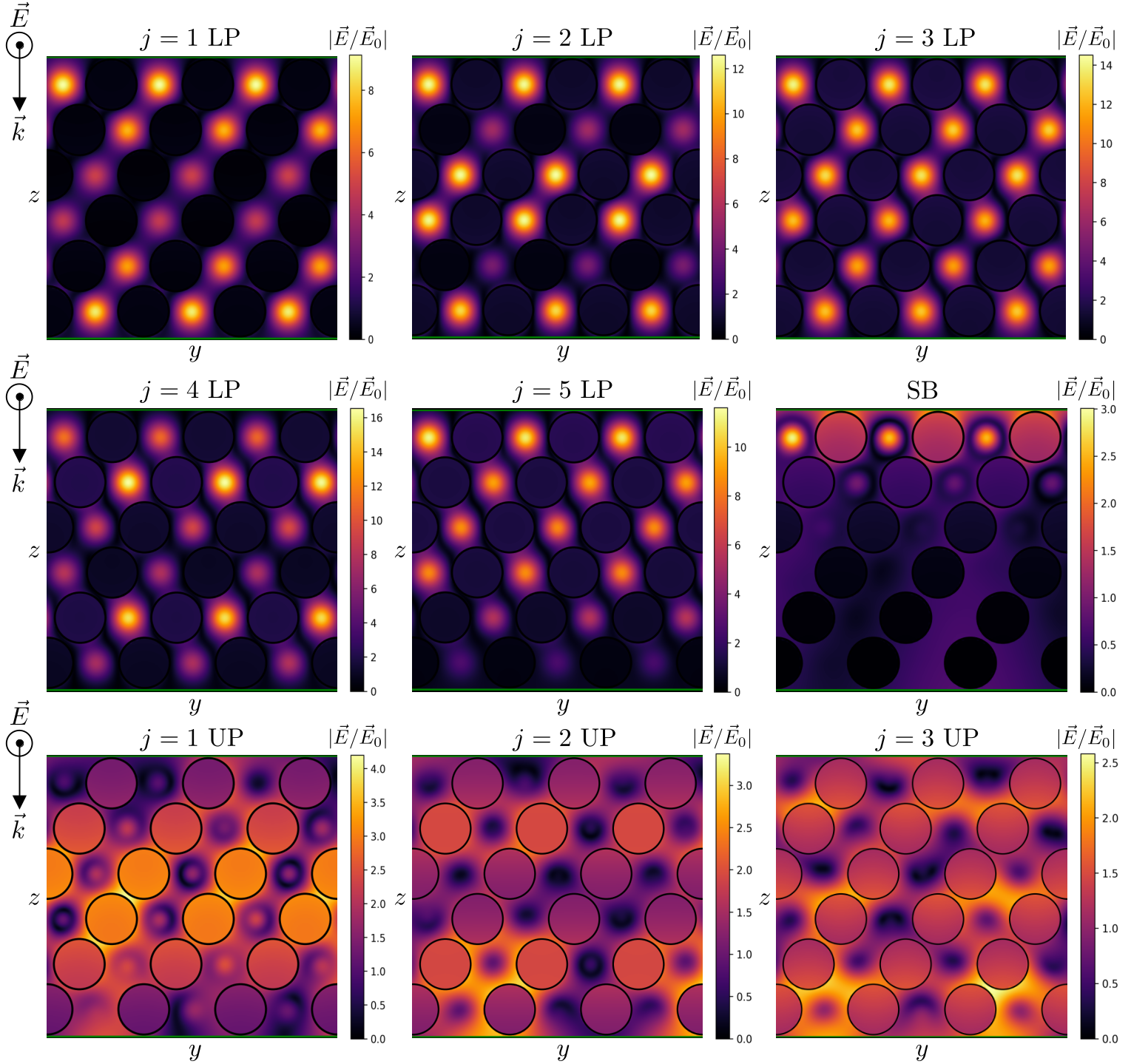}
    \caption{Near-field maps for perfectly ordered six-layered Au NP films at all observed $j$ modes corresponding to both lower polariton (LP) and upper polariton (UP) branches and at one frequency inside the stop band (SB). The presented field maps are computed on the $yz$ plane that contains $\vec{k}$, but $\vec{E}$, in this case, is orthogonal to the plane. The direction of the $\vec{E}$ corresponds to the EX configuration.}
    \label{SIfig:NF_TE_xz_ASL}
\end{figure}

\begin{figure}[htbp!]
    \centering
    \includegraphics[width=1.0\textwidth]{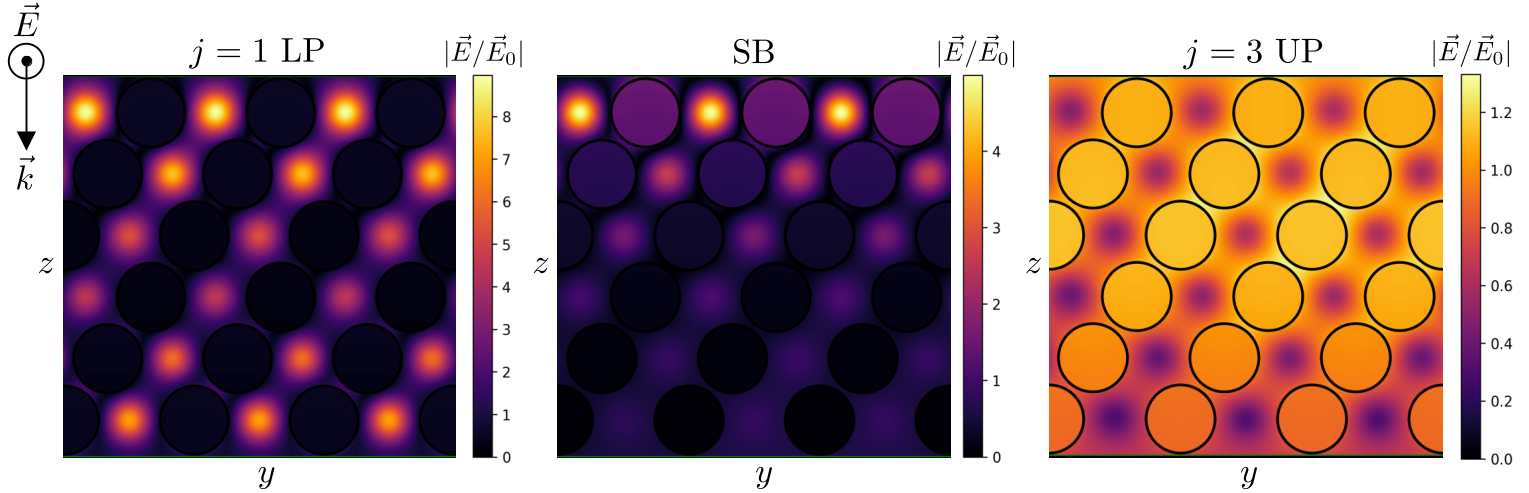}
    \caption{Near-field maps for perfectly ordered six-layered ITO NP films at all observed $j$ modes corresponding to both lower polariton (LP) and upper polariton (UP) branches and at one frequency inside the stop band (SB). The presented field maps are computed on the $yz$ plane that contains $\vec{k}$, but $\vec{E}$, in this case, is orthogonal to the plane. The direction of the $\vec{E}$ corresponds to the EX configuration.}
    \label{SIfig:NF_TE_xz_ISL}
\end{figure}

\begin{figure}[htbp!]
    \centering
    \includegraphics[width=1.0\textwidth]{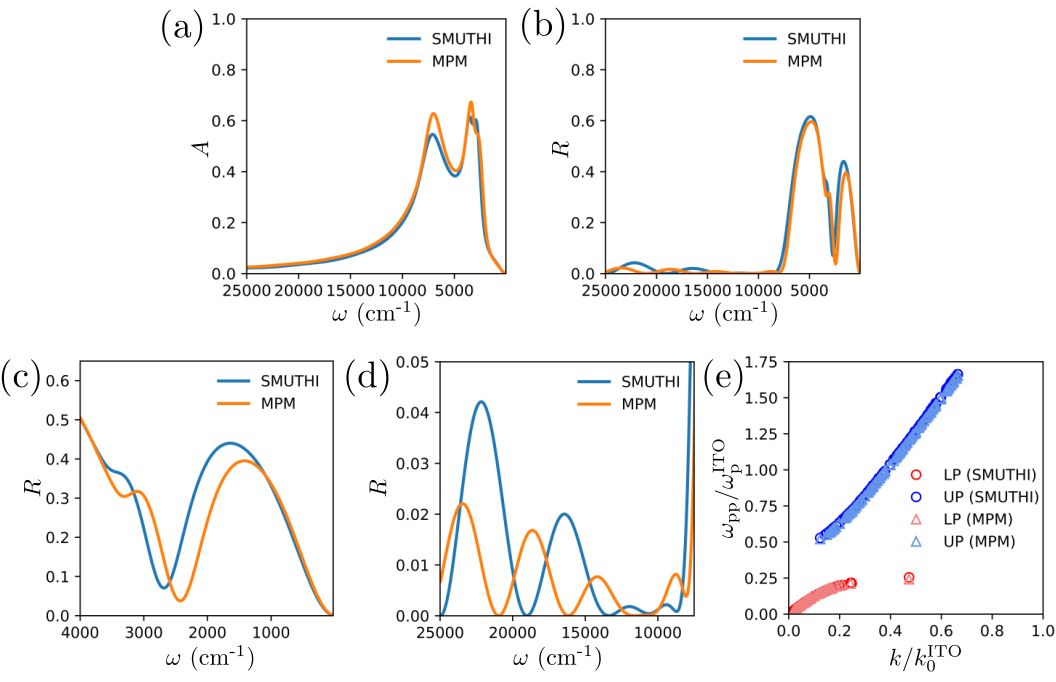}
    \caption{Comparisons between SMUTHI and MPM results for fifty-layered perfectly ordered ITO SLs: (a) Absorptance and (b) reflectance spectra, (c) zoomed-in lower and (d) upper polariton branches of the reflectance spectra shown in (b), and (e) dispersion data sets. `LP' and `UP' stand for lower and upper polariton, respectively.}
    \label{SIfig:MPM_Smuthi_comp}
\end{figure}

\FloatBarrier

\setcounter{table}{0}
\renewcommand{\tablename}{Table}
\renewcommand{\thetable}{S\arabic{table}}

\section{Supporting Tables}

\begin{table}[htbp!]
\centering
  \begin{tabular}{|>{\centering\arraybackslash}m{1.0in}   >{\centering\arraybackslash}m{1.0in}  >{\centering\arraybackslash}m{1.0in}  >{\centering\arraybackslash}m{1.0in}  >{\centering\arraybackslash}m{1.0in} |}
    \hline
     \textbf{$t_{2d}$} & \textbf{$t_{3d}$} & \textbf{$\psi_6$} & \textbf{$q_6$} & \textbf{$q_6/q_{6,FCC}$}\\ [0.5ex] 
    \hline
0.1 & 0.1 & 0.65 & 0.45 & 0.77 \\ \hline
0.2 & 0.1 & 0.71 & 0.46 & 0.79 \\ \hline
0.3 & 0.1 & 0.71 & 0.45 & 0.78 \\ \hline
0.5 & 0.1 & 0.74 & 0.46 & 0.80 \\ \hline
1 & 0.1 & 0.81 & 0.48 & 0.83 \\ \hline
10 & 0.1 & 0.86 & 0.49 & 0.84 \\ \hline
100 & 0.1 & 0.89 & 0.50 & 0.86 \\ \hline
200 & 0.1 & 0.92 & 0.51 & 0.87 \\ \hline
 \end{tabular}
 \caption[Brownian Dynamics Simulation Parameters]{Simulation parameters and calculated order parameters of Brownian dynamics configurations used in the main text. $t_{2d}$ is the time to compress each monolayer in 2D, and $t_{3d}$ is the time to compress the stack of monolayers in 3D, both in units of diffusion time $\tau_D$.}
\label{table:pol_bd_sim}
\end{table}

\begin{table}[htbp!]
    \centering
    \begin{tabular}{|l c c c c|}
    \hline
     & $n_{\rm eff}$ & $\omega_{\rm CPR}$ & $\Omega_{\rm R}$ & $\eta$ \\
    \hline
    Au films & \multirow{2}{*}{$1.23$} & $20.8 \times 10^3$ cm\textsuperscript{-1} & $23.1 \times 10^3$ cm\textsuperscript{-1} & \multirow{2}{*}{1.11} \\
    ($l=1, m=1$) & & ($\simeq 2.58$ eV) & ($\simeq 2.86$ eV) & \\
    \hline
    ITO films & \multirow{2}{*}{$1.70$}  & $3.83 \times 10^3$ cm\textsuperscript{-1} & $ 2.77 \times 10^3$ cm\textsuperscript{-1} & \multirow{2}{*}{0.72} \\
    ($l=1, m=1$) & & ($\simeq 0.47$ eV) & ($\simeq 0.34$ eV) & \\
    \hline
    ITO films & \multirow{2}{*}{$1.72$} & $3.88 \times 10^3$ cm\textsuperscript{-1} & $2.84 \times 10^3$ cm\textsuperscript{-1} & \multirow{2}{*}{$0.73$} \\
    ($l=2, m=2$) & & ($\simeq 0.48$ eV) & ($\simeq 0.35$ eV) & \\
    \hline
    ITO films & \multirow{2}{*}{$1.70$} & $3.84 \times 10^3$ cm\textsuperscript{-1} & $2.74 \times 10^3$ cm\textsuperscript{-1} & \multirow{2}{*}{$0.71$} \\
    (MPM) & & ($\simeq 0.48$ eV) & ($\simeq 0.34$ eV) & \\
    \hline
    \end{tabular}
    \caption{Dispersion fitting parameters and the normalized coupling strength $\eta$ for both Au and ITO films.}
    \label{table:compare_fit_params}
\end{table}

\begin{table}[htbp!]
    \centering
    \begin{tabular}{|m{2cm} m{3cm} m{3cm}|}
    \hline
    Quantities                      & Au NP films   & ITO NP films  \\
    \hline
    $\epsilon_{\infty}$             & 1.0             & 4.0             \\
    $f(k)$                          & -1.01           & -1.37         \\
    $\eta_{\rm Th}$                 & 1.006           & 0.74          \\
    $\tilde{\eta}$                  & 1.34            & 1.08         \\
    \hline
    \end{tabular}
    \caption{Quantities related to the theoretical estimation of the normalized coupling strength, $\eta_{\rm Th}$, for both Au and ITO NP films considering $\epsilon_{\infty}$ instead of $\epsilon_{d}$ in the calculations of $\eta_{\rm Th}$ and $\tilde{\eta}$.}
    \label{table:compare_w_eps_inf}
\end{table}

\end{document}